\documentclass[a4paper,12pt,twoside,epsf]{book}
\usepackage[latin1]{inputenc}
\usepackage[english,UKenglish]{babel}
\usepackage{fontenc}
\usepackage{graphicx}
\usepackage{multirow}
\usepackage{setspace}
\usepackage{amssymb}
\usepackage{array}
\usepackage{amsmath}
\usepackage{booktabs}
\usepackage{tabularx}
\usepackage{enumerate}
\usepackage[active]{srcltx}

\def\ebh{e^{-\beta_n \hat H}}
\def\intinf{\int_{-\infty}^{\infty}}
\def\ootph{\frac{1}{2\pi\hbar}}
\begin{document}
\bibliographystyle{tim1}
\setcounter{secnumdepth}{4}
\thispagestyle{empty}
\begin{center}
\vspace*{5mm}
{\LARGE \bf 
{An Electronically Non-Adiabatic}

{Generalization of}

{Ring Polymer Molecular Dynamics}

}
\vspace*{10mm}

{\large \rm

{Timothy J.\ H.\ Hele}

{Exeter College,} {Oxford University}

}
\newlength{\figwidthc}
\setlength{\figwidthc}{\columnwidth}
\begin{figure}[h]
\centering
\resizebox{\figwidthc}{!} {\includegraphics[angle = 270]{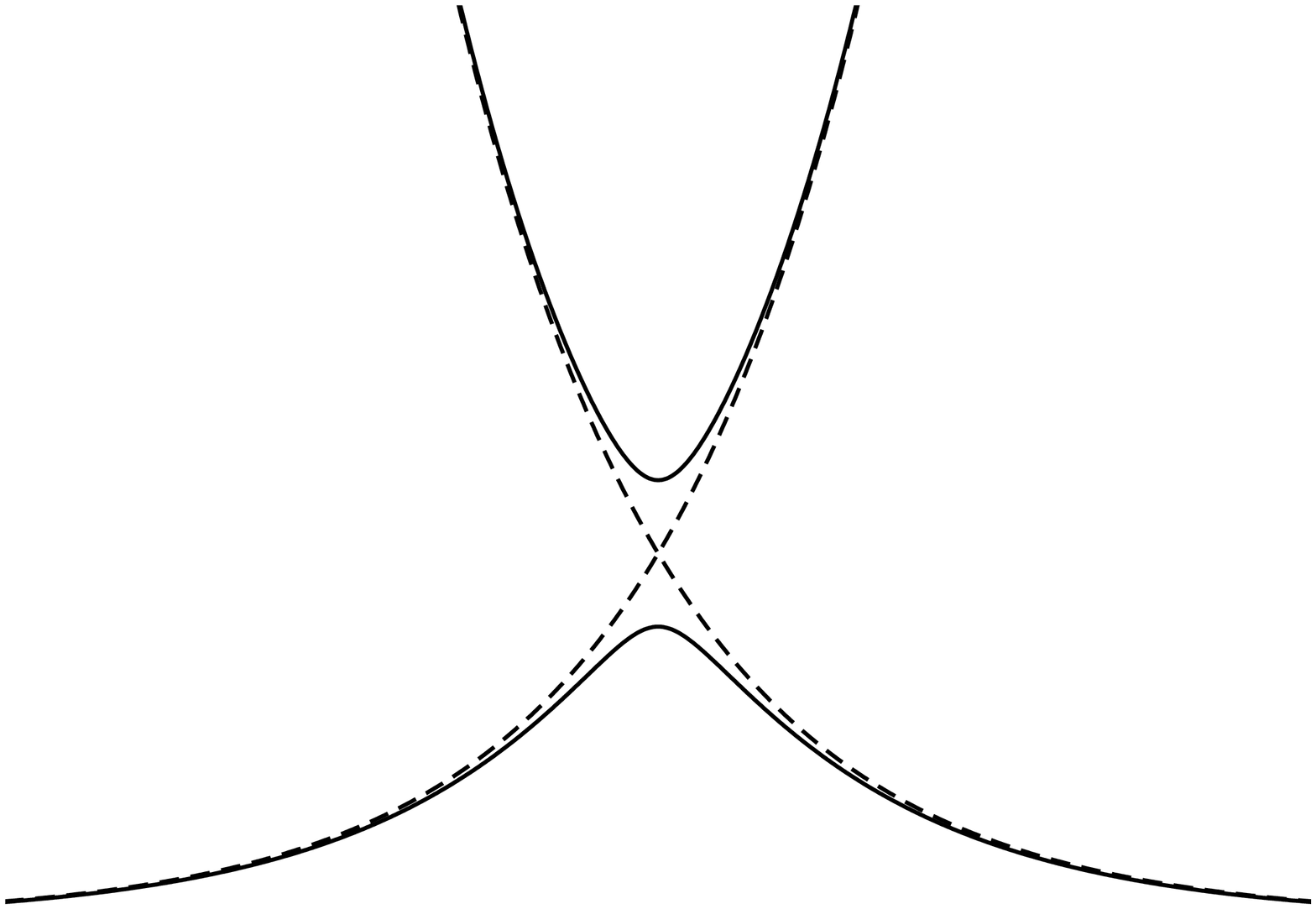}}
\end{figure}
{\large \it
A thesis submitted for the
Honour School of Chemistry:

Chemistry Part II

June 2011
}
\end{center}
\section*{Summary}
\pagenumbering{roman}
\setcounter{page}{2}
Electronically non-adiabatic effects, in which the nuclear motion occurs on more than one electronic potential energy surface, are of fundamental importance to numerous branches of chemistry and biology, ranging from quantum dissipation to photosynthesis. In this thesis I generalize Ring Polymer Molecular Dynamics (RPMD) rate theory to electronically non-adiabatic systems, followed by application to two one-dimensional curve crossing models and a multidimensional spin-boson model.

RPMD rate theory uses the classical isomorphism to model a quantum mechanical system as many copies of a classical system connected by harmonic springs, forming a ring polymer necklace. By evolving the ring polymer with classical molecular dynamics through an extended classical phase space, the RPMD reaction rate can be calculated. The RPMD rate captures quantum mechanical effects, is independent of the choice of the location of the dividing surface between products and reactants, and reduces to classical rate theory in the high temperature limit. Here I extend RPMD rate theory to include electronically non-adiabatic effects, leading to a new expression for the potential energy which the ring polymer experiences. 

Numerous desirable features of the non-adiabatic ring polymer potential are illustrated, such as physically plausible results in the high temperature limit for both widely separated and degenerate electronic states. In the limit of a single electronic potential energy surface the conventional adiabatic RPMD potential is recovered, and for a system whose Hamiltonian can be separated into electronically coupled and electronically independent components, only the electronically coupled part need be considered using the more computationally demanding non-adiabatic ring polymer potential.

The generalisation of RPMD rate theory to non-adiabatic systems is then applied to two one-dimensional Landau-Zener curve crossing models, initially a symmetric model where the optimum dividing surface is evident by symmetry, and then an asymmetric model where the optimum dividing surface is calculated variationally. Thermal RPMD reaction rates and classical rates are calculated over a wide temperature range for both the electronically coupled model, and the adiabatic model where the reactant moves solely on the lower potential energy surface. Exact quantum rates are also available for comparison. The RPMD reaction rates are found to agree with exact quantum results roughly to within a factor of two in the symmetric model, even in deep tunnelling at 100 K, where the classical rate is in error by many orders of magnitude. For the asymmetric model, RPMD rate theory is an even better approximation to the exact quantum rate, and in both cases the departure of the RPMD rate from the quantum rate in the electronically coupled system is comparable to that in the adiabatic system.

The spin-boson model is then considered, which consists of a spin system bilinearly coupled to a harmonic bath. It has applications ranging from electron transfer in solution to quantum entanglement, and a variety of quantum exact and approximate reaction rates are available for comparison. In this thesis the bath is defined by the Debye spectral density, whose long frequency tail leads to the mass of the reaction co-ordinate vanishing in the limit of an infinite number of bath modes. 

The non-adiabatic form of RPMD rate theory is extended to multidimensional systems, and the Debye spectral density is found to present numerous challenges, such as the Quantum Transition State Theory rate being undefined. 
However, by an unusual factorisation of the RPMD reaction rate into the probability density of ring polymer centroids at the dividing surface and the flux of positive-momentum centroids which are found in the product region at long time, the RPMD rate is calculable as the product of two convergent factors. Furthermore, the RPMD reaction rate is found to be closer to the exact quantum rate than almost all other approximate methods which are available for comparison.

These results suggest a new methodology for rate calculation of electronically non-adiabatic systems, whose computational effort scales linearly with system size, but includes an accurate modelling of recrossing dynamics and quantum effects.

\tableofcontents
\section*{Acknowledgements}

I am indebted to my supervisor, David Manolopoulos, for his insight and instruction this year. I am also grateful to Guy, Michele and Yury for their assistance, and for the many Aarts and Logan group members who have provided encouragement and amusement.

Lastly I wish to thank my family and friends.

\chapter{Introduction}
\label{chap:intro}
\pagenumbering{arabic}
\setcounter{page}{1}
Electronically non-adiabatic effects are of fundamental importance to numerous processes in chemistry and biology. Their diverse applicability includes spectroscopy, pericyclic reactions, photochemistry and reaction kinetics \cite{photochem, surf_chem, spec, na_rate, sb, sbprev, biopt}.
This study concerns the extension of Ring Polymer Molecular Dynamics (RPMD) rate theory to non-adiabatic systems, with application initially to two one-dimensional barrier transmission models, and then to a multidimensional spin-boson model.

RPMD rate theory uses the classical isomorphism \cite{miller_isomorphism} to model quantum mechanical systems in an extended classical phase space \cite{Mano_cen}, whereby multiple copies of the classical system are connected by harmonic springs. RPMD rate theory takes this phase space literally, and uses its real-time dynamics to evaluate chemical reaction rates. Because of its classical nature, the theory is comparatively simple to implement, even for systems of high dimensionality where exact quantum calculations are prohibitively expensive. For adiabatic reactions the theory is exact in the high-temperature limit, exact for a parabolic barrier \cite{Mano_pin} and independent of the location that is used for the dividing surface between reactants and products \cite{Mano_cen}. This last characteristic is particularly important for complex systems where calculating the location of the optimum dividing surface is highly non-trivial. Its versatility has lead to applications both in condensed media \cite{proton_transfer, h_mu_diffusion} and gas phase reactions \cite{CH_4, bi_rate}.

Furthermore, RPMD theory has been shown to be surprisingly accurate, even in deep tunnelling where purely classical theories are in error by many orders of magnitude \cite{h_mu_diffusion, bi_rate}. A recent study by Richardson and Althorpe \cite{althorpe} has shown how, in deep tunnelling, there is a connection between RPMD rate theory and the semi-classical instanton model \cite{semiclas_in}, thereby explaining the success of RPMD in this regime. They have also explained, for adiabatic reactions at low temperatures, why the RPMD rate is underestimated for symmetric barriers and overestimated for asymmetric barriers compared to exact quantum results.

The aim of this thesis is to extend RPMD rate theory to non-adiabatic reactions, where the reagents move on more than one electronic potential energy surface. The motivation for this application is the wide variety of chemical and biological processes which are inherently non-adiabatic, ranging from photochemical pericyclic reactions to surface chemistry \cite{photochem, surf_chem}. For example, the F($^2$P) + H$_2$ reaction proceeds on three coupled potential energy surfaces arising from the $\rm ^2P_{3/2}$ and $\rm ^2P_{1/2}$ levels of the fluorine atom \cite{F2}, while photosynthesis involves a sequence of electron transfers between protein complexes leading to the eventual reduction of CO$_2$ to form glucose \cite{photo1, photo2}.

In order to test the multidimensional validity of RPMD rate theory for non-adiabatic processes, it is applied to a spin-boson model, consisting of a spin degree of freedom bilinearly coupled to a harmonic bath. This model of quantum dissipation has applications ranging from biological electron transfer \cite{garg}, to describing the entanglement between a qubit and its environment in a quantum computer \cite{qcomp}.

RPMD theory is derived in chapter~\ref{chap:rpmd} and its previous applications are outlined. In chapter~\ref{chap:nag} the theory is extended to non-adiabatic systems, followed by an application to two simple Landau-Zener curve crossing models in chapter~\ref{chap:1D}. A spin-boson model with Debye spectral density is then examined in chapter~\ref{chap:sb}, followed by a brief conclusion in chapter~\ref{chap:c}.
\chapter{Ring Polymer Rate Theory}
\label{chap:rpmd}
In this chapter a variety of methods for calculating reaction rates is explored. The exact quantum rate is initially considered, followed by classical rate theory and classical transition state theory. The isomorphism between a classical path integral and a quantum mechanical partition function is derived in section \ref{sec:isomorphism}, from which Ring Polymer Molecular Dynamics (RPMD) rate theory arises. Quantum transition state theory (QTST) is then introduced as a limiting case of RPMD rate theory, and the chapter concludes with a discussion of previous applications. For simplicity, a one-dimensional barrier transmission problem is used to illustrate the theory, although (as will be shown in section \ref{sec:multidimen}) the theory can easily be generalized to more complex systems.

\section{Quantum mechanical rate theory}
\label{sec:QMratetheory}
For a one-dimensional system with a Hamiltonian 
\begin{equation}
\hat H = \hat T + \hat V = \frac{\hat p^2}{2m} + V(\hat q),
\label{eq:ham}
\end{equation}
where $V(\hat q)$ is similar to that illustrated in Fig~\ref{fig:v}, the exact quantum mechanical reaction rate is \cite{miller1, miller2}
\newlength{\figwidthv}
\setlength{\figwidthv}{\columnwidth}
\begin{figure}[h]
\centering
\resizebox{\figwidthv}{!} {\includegraphics[angle = 270]{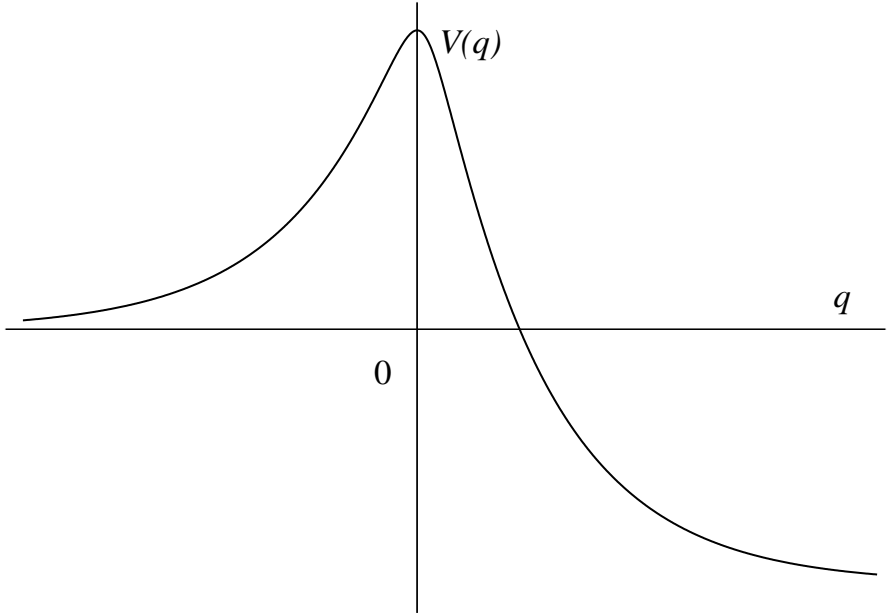}}
\caption{Illustrative one-dimensional potential for a barrier transmission problem. $V(q \to - \infty) = 0$, there is a well-defined barrier, and $V(q \to + \infty) =$ const.}
\label{fig:v}
\end{figure}
\begin{equation}
k(T) = \frac{1}{Q_r(T)} \lim_{t \rightarrow \infty} \tilde c_{fs}(t),
\label{eq:kT}
\end{equation}
where $k(T)$ is the rate at temperature $T$ and $Q_r(T)$ is the reactant partition function per unit length. The Kubo-transformed flux-side correlation function, $\tilde c_{fs}(t)$, is defined as \cite{Kubo}
\begin{equation}
\tilde c_{fs}(t) = \frac{1}{\beta} \int_0^{\beta} {\rm Tr}\left[ e^{-(\beta - \lambda) \hat H} \hat F e^{- \lambda \hat H} e^{+i \hat H t/ \hbar} \hat h e^{-i \hat H t/\hbar} \right]d\lambda.
\label{eq:cfs_quantum}
\end{equation}
Here $ \beta \equiv 1/k_BT$, $k_B$ is the Boltzmann constant and $\hat H$ is the Hamiltonian in Eq~(\ref{eq:ham}). $\hat F$ is the flux operator, which measures the rate of passage through the dividing surface $q^{\ddag}$ along the reaction co-ordinate $q$. Quantum mechanically, this is expressed as the Heisenberg time derivative of the side operator $\hat h$, which projects onto states which are on the product side of the dividing surface, such that 
\begin{equation}
\hat F \equiv \frac {i}{\hbar} \left[\hat H, \hat h \right],
\end{equation}
with 
\begin{equation}
 \hat h \equiv h(\hat q - q^{\ddag}),
\end{equation}
where 
\begin{equation}
h(q - q^{\ddag}) = 
\left\{ \begin{array}{cc}
	0, & q < q^{\ddag}  \\
	1, & q > q^{\ddag} \\
	\end{array}\right..
\end{equation}

A qualitative understanding of the flux-side correlation function can be gained by considering the contents of the trace in Eq~(\ref{eq:cfs_quantum}) as a time-evolved side operator and a Boltzmann-weighted flux operator. The time-evolved side operator projects onto configurations which, in the long time limit, are on the product side of the dividing surface. For these configurations, the flux operator measures their rate of passage through the dividing surface at time $t = 0$, and the Boltzmann operator (here `smeared' by an integral over $\lambda$) ensures a thermal rate is produced. 

There exist other flux-side correlation functions which give the same exact rate in the long time limit, but differ in their treatment of the Boltzmann operator. These include no splitting \cite{miller1}
\begin{equation}
c_{fs}^{ns}(t) = {\rm Tr}\left[ e^{-\beta\hat H} \hat F e^{+i \hat H t/ \hbar} \hat h e^{-i \hat H t/\hbar} \right],
\end{equation} 
and symmetric splitting \cite{miller2}
\begin{equation}
c_{fs}^{sym}(t) = {\rm Tr}\left[ e^{-\beta\hat H/2} \hat F e^{- \beta\hat H/2} e^{+i \hat H t/ \hbar} \hat h e^{-i \hat H t/\hbar} \right],
\end{equation}
amongst others explored in Ref.~\cite{miller2}. All of these functions are independent of the location of the dividing surface in the long time limit, a consequence of the quantum mechanical continuity equation \cite{Mano_cen}. The advantage of using the so-called `Kubo-transformed' correlation function in Eq~(\ref{eq:cfs_quantum}) is its relation to the RPMD rate formula \cite{Mano_pin}, and that it is a real and odd function of time \cite{Mano1}. It is also a continuous function of time, meaning that one cannot define a rigorous `Quantum Transition State Theory' rate, as the $t \rightarrow 0_+$ limit of the exact quantum mechanical reaction rate, $\tilde c_{fs}(t \rightarrow 0) = 0$ \cite{Mano_pin}.

\section{Classical rate theory}
The rate coefficient is now defined as the long time limit of the {\it classical} flux-side correlation function,
\begin{equation}
k^{cl}(T) = \frac{1}{Q_r(T)} \lim_{t \rightarrow \infty} c_{fs}^{cl}(t).
\label{eq:clas_k}
\end{equation}
To form a classical correlation function from a quantum one, quantum mechanical operators are converted into their classical counterparts which commute, and the trace over states becomes an integral over the classical phase space of positions and momenta. The flux operator, which cannot be evaluated directly from the configuration of the system, is now expressed as the classical time derivative of the side operator,
\begin{equation}
F = \frac{d}{dt}h(q-q^{\ddag}) = \delta(q-q^{\ddag}) \frac{dq}{dt}=\delta(q-q^{\ddag})\frac{p}{m},
\end{equation}
where $\delta(q-q^{\ddag})$ is a Dirac delta function, $p$ is the momentum and $m$ the mass. Consequently,
\begin{equation}
c_{fs}^{cl}(t) = \frac{1}{2\pi\hbar} \int dp_0 \int dq_0\ e^{-\beta H(p_0,q_0)}\delta(q_0-q^{\ddag})\frac{p_0}{m}h(q_t-q^{\ddag}).
\label{cfs_clas}
\end{equation}
Here, $p_0$ and $q_0$ refer to the momentum and position at zero time, and $H(p_0,q_0)$ is the classical Hamiltonian of the system. This Hamiltonian can be used to evolve the system to time $t$ by conventional molecular dynamics, thereby avoiding the quantum time evolution operators $e^{\pm i \hat H t/ \hbar}$ which present a greater computational challenge \cite{sbprev}. 

The physical interpretation of the classical flux-side correlation function in Eq~(\ref{cfs_clas}) is similar to that of the exact quantum mechanical correlation function, except for the portrayal of the flux operator as a Dirac delta multiplied by an initial velocity, which implies that a starting configuration is chosen from a Boltzmann distribution that is initially constrained to the dividing surface. The system is evolved to long time, after which the trajectory will be on one side of the barrier or another, and the position noted. If the position is in the product region, the side function \mbox{$h(q-q^{\ddag})=1$} and the initial velocity $v_0 = p_0/m$ contributes to the flux-side correlation function. If the particle resides in the reactant region, its initial velocity $v_0$ does not contribute to the overall rate.

The classical rate, like the exact quantum rate, is independent of the location of the dividing surface, in this case owing to Liouville's theorem \cite{miller1, Mano_notes}. By its classical nature, no quantum effects such as tunnelling are included in the classical rate calculation and large deviations (many orders of magnitude) from the exact quantum rate are commonplace at low temperatures where quantum effects are particularly pronounced \cite{bi_rate}.

\section{Classical transition state theory}
\label{sec:cltst}
The evaluation of the classical rate in Eq~(\ref{eq:clas_k}) involves a long time limit, which, for complex systems, may prove expensive to calculate. In this section the \mbox{$t \rightarrow 0_+$} limit is explored, where only the behaviour at the transition state, $q = q^{\ddag}$, is required. In the exact quantum case (section \ref{sec:QMratetheory}), the short-time limit of the flux-side correlation function was zero, but in the classical case a finite rate is defined. Whether the reactant will be on the product side of the dividing surface or not, in the infinitesimal time limit, is simply a question of whether the initial momentum is positive or negative, such that $\lim_{t \to 0+} h(q_t - q^{\ddag}) = h(p_0)$. Consequently, the integral in Eq~(\ref{cfs_clas}) can be separated into one over momenta and one over positions:
\begin{eqnarray}
k^{cl, TST}(T) & = & \frac{c_{fs}^{cl}(t \rightarrow 0_+)}{Q_r(T)} \label{eq:k_clas}\\
& = & \frac{1}{2\pi\hbar Q_r(T)} \int dp_0 \int dq_0\ e^{-\beta H(p_0,q_0)}\delta(q_0-q^{\ddag})\frac{p_0}{m}h(p_0)  \nonumber \\
& = & \frac{1}{2\pi\hbar Q_r(T)} \int dp_0\  e^{-\beta p_0^2/2m}\frac{p_0}{m}h(p_0)\int dq_0\ e^{-\beta V(q_0)}\delta (q_0-q^{\ddag}) \nonumber \\
& = & \sqrt{\frac{\beta}{2\pi m}} \int dp_0\ e^{-\beta p_0^2/2m}\frac{p_0}{m}h(p_0) e^{-\beta V(q^{\ddag})} \nonumber \\
& = & \frac{1}{2} \langle | \dot q | \rangle e^{-\beta V(q^{\ddag})}.
\label{clas_tst}
\end{eqnarray}
We have noted that, for a barrier transmission model, the partition function per unit length is the inverse de Broglie thermal wavelength,
\begin{equation}
Q_r(T) = \frac{1}{\Lambda(T)} = \sqrt{\frac{m}{2\pi\beta\hbar^2}},
\label{eq:debroglie1}
\end{equation}
and the mean magnitude of the velocity is defined as
\begin{eqnarray}
\langle | \dot q | \rangle & = & \frac{\int_{-\infty}^{\infty} dp \ e^{-\beta p^2/2m}\frac{|p|}{m}}{\int_{-\infty}^{\infty} dp \ e^{-\beta p^2/2m}}. 
\end{eqnarray} 
The result in Eq~(\ref{clas_tst}) can be interpreted as the flux through the dividing surface, halved to account only for reactant $\to$ product motion, multiplied by the Boltzmann probability that the reagent will reach the top of the barrier.

Classical transition state theory (classical TST) has the advantage of ease of calculability, as it does not require any molecular dynamics. However, Eq~({\ref{clas_tst}) suffers the major disadvantage of being exponentially sensitive to the location of the dividing surface; from Eq~(\ref{clas_tst}),
\begin{equation}
k^{cl, TST}(T) \propto e^{-\beta V(q^{\ddag})}.
\label{eq:exp_sen}
\end{equation}
For a one-dimensional barrier transmission model, the classical TST rate will always be equal to or greater than the classical rate, and equal to the classical rate for the optimum dividing surface (where $e^{-\beta V(q^{\ddag})}$ is at a minimum, and no recrossings occur) \cite{varopt}. As such, the optimum dividing surface may be chosen variationally. For models of higher dimensionality, the classical TST rate with the optimum dividing surface may still deviate from the classical rate theory result due to the neglect of barrier recrossing. 

\section{The Classical Isomorphism}
\label{sec:isomorphism}
It has been known since the 1980s \cite{miller_isomorphism} that static equilibrium properties of a quantum mechanical system (such as energy, entropy etc.) can be calculated by an integral over a fictitious ring polymer in an extended classical phase space. Here I sketch how this arises for a single electronic potential, in order to prepare for the generalization to an electronically coupled model in chapter~\ref{chap:nag}. For a more detailed discussion of path integrals, the reader is referred to the substantial literature on the subject \cite{miller_isomorphism, feynman_path_integrals, feynman_stat_mech, schulman, PI_review}. In a later section this isomorphism is applied to real time correlation functions, leading to RPMD rate theory.

A quantum mechanical partition function, $Z$, can be expressed as 
\begin{equation}
Z = {\rm Tr}[e^{-\beta \hat H}],
\end{equation}
where $\hat H$ is the Hamiltonian of the system in Eq~(\ref{eq:ham}). This partition function can be evaluated using the Trotter discretization \cite{schulman},
\begin{equation}
Z = \lim_{n \rightarrow \infty} Z_n = \lim_{n \rightarrow \infty} {\rm Tr} [(\ebh)^n],
\label{eq:trotter}
\end{equation}
where $\beta_n = \beta/n$, with $n$ being the number of fictitious ring polymer beads. One can approximate (to second order in $\beta_n$) that 
\begin{equation} 
\ebh \simeq e^{-\beta_n \hat V /2} \ e^{-\beta_n \hat T} \ e^{-\beta_n \hat V /2},
\label{split} 
\end{equation}
such that
\begin{equation}
 Z_n = {\rm Tr} [(\ebh)^n] \simeq {\rm Tr} \left[\left(e^{-\beta_n \hat V /2} \ e^{-\beta_n \hat T} \ e^{-\beta_n \hat V /2}\right)^n\right].
\end{equation}
Noting the possibility of cyclic permutation within a trace, one can move the final $e^{-\beta_n \hat V /2}$ in the trace to the front without any loss of accuracy, forming
\begin{equation}
 Z_n = {\rm Tr} \left[\left(e^{-\beta_n \hat V} \ e^{-\beta_n \hat T} \right)^n\right].
\label{eq:symsplit}
\end{equation}
The trace can be evaluated in any basis, so choosing the co-ordinate representation,
\begin{equation}
Z_n = \int dq_1  \langle q_1 |\left(e^{-\beta_n \hat V} \ e^{-\beta_n \hat T} \right)^n| q_1 \rangle.
\end{equation}
Inserting the unit operator $ \hat 1 = \int dq_i | q_i \rangle \langle q_i |$ for $i = 2,3, \ldots, n$ between each $e^{-\beta_n \hat V} \ e^{-\beta_n \hat T}$ term,
\begin{equation}
 Z_n = \int dq_1 \ldots \int dq_n \langle q_1 | e^{-\beta_n \hat V} \ e^{-\beta_n \hat T} | q_2 \rangle \ldots \langle q_n | e^{-\beta_n \hat V} \ e^{-\beta_n \hat T} | q_1 \rangle,
\label{qstates}
\end{equation}
leading to an integral over $n$ similar terms. As the potential energy is diagonal in the co-ordinate representation, i.e.\ $\hat V |q \rangle = V(q)|q\rangle$,
\begin{equation}
\langle q | e^{-\beta_n \hat V} \ e^{-\beta_n \hat T} | q' \rangle \simeq e^{-\beta_n V(q)} \ \langle q | e^{- \beta_n \hat T}  | q' \rangle.
\label{eq:qq'}
\end{equation}
Inserting the momentum unit operator, $ \hat 1 = \int dp | p \rangle \langle p |$, and noting that $\hat T |p\rangle = \frac {p^2}{2m}|p\rangle$,
\begin{equation}
\langle q| e^{-\beta_n \hat T}| q' \rangle = \int dp \ \langle q|p \rangle e^{-\beta_n p^2 / 2m} \langle p|q'\rangle.
\label{aa}
\end{equation}
However, $\langle q|p\rangle$ is the momentum eigenstate $|p\rangle$ in the position representation of a free particle \cite{Mano_notes, Dirac},
\begin{equation}
 \langle q | p \rangle = \frac{1}{\sqrt{2 \pi \hbar}} e^{-ipq/\hbar}.
\end{equation}
Utilising this in Eq~(\ref{aa}), and then completing the square with the substitution $p' = p - im(q-q')/\beta_n \hbar$,
\begin{eqnarray}
\langle q| e^{-\beta_n \hat T}| q' \rangle & = & \ootph \intinf dp \ e^{- \beta_n p^2/2m + ip(q-q')/\hbar} \label{eq:pqint} \\
& = & \ootph \left( \frac{2 \pi m}{\beta_n}\right)^{1/2} \ e^{- \frac{1}{2}\beta_n m \omega_n^2 (q - q')^2},
\label{eq:qTq}
\end{eqnarray}
where $\omega_n = 1/\beta_n \hbar$. Finally, noting that
\begin{equation} 
\intinf dp \ e^{- \beta_n p^2 /2m} = \left(\frac{2 \pi m}{\beta_n}\right)^{1/2}, 
\end{equation}
one can reinsert momentum states into Eq~(\ref{eq:qTq}) \cite{parrinello}. Considering the bra-ket involving the full Hamiltonian, $ \hat H$, rather than just the kinetic energy operator $\hat T$, this gives
\begin{equation}
 \langle q| e^{-\beta_n \hat H}| q' \rangle = \ootph \intinf dp \ e^{-\beta_n[p^2/2m + \frac{1}{2}m\omega_n^2(q-q')^2 + V(q)]}.
\end{equation}
The partition function can therefore be expressed as 
\begin{equation}
Z_n = \frac{1}{(2\pi\hbar)^n} \int dp_1 \ldots \int dp_n \int dq_1 \ldots \int dq_n \ e^{-\beta_n H_n({\bf p,q})},
\label{eq:partition_function}
\end{equation}
where 
\begin{equation}
 H_n({\bf p,q}) = \sum_{j = 1}^{n} \left[\frac{p_j^2}{2m} + \frac{1}{2}m\omega_n^2 (q_j - q_{j+1})^2 + V(q_j)\right],
\label{eq:hamiltonian}
\end{equation}
subject to the cyclic boundary condition $q_{n+1} \equiv q_1 $. Here $\bf p$ is the vector of bead momenta $\{p_j\}$, and likewise $\bf q$ of bead positions. We see that in the case of a single electronic potential energy surface, the potential arising from $\hat V$ is a simple additive sum of contributions from all beads,
\begin{equation}
 V_n({\bf q}) = \sum_{j=1}^n V(q_j).
\label{eq:adv}
\end{equation}
Lastly, the multiple integrals in Eq~(\ref{eq:partition_function}) are commonly rewritten in terms of $\bf p$ and $\bf q$, leading to the neater expression
\begin{equation}
 Z_n = \frac{1}{(2\pi\hbar)^n} \int d {\bf p} \int d {\bf q}\ e^{- \beta_n H_n ({\bf p,q})}.
\label{eq:Z}
\end{equation}
The only approximation in deriving Eq~(\ref{eq:Z}) is forming the symmetric split operator in Eq~(\ref{split}), which has local error of $O(\beta_n^3)$. As there are $n$ beads present the global error scales as $O(\beta_n^2)$. Consequently, in the limit as $n \rightarrow \infty $ and $\beta_n \to 0$ the partition function becomes exact (at any finite temperature).

The physical interpretation of Eq~(\ref{eq:Z}) is that the quantum mechanical partition function of a single particle can be expressed as the classical partition function of a necklace of beads, each with the mass of the original quantum particle and with harmonic springs between neighbouring beads, each with the same spring constant $m \omega_n^2 \propto (nT)^2$. The extended system exists at a virtual temperature of $nT$, and in the high temperature limit, the spring constant stiffens until the ring polymer collapses into a single, classical, particle.

At lower temperatures, when the spring constant slackens, the radius of gyration of the ring polymer $r_G$ increases and the ring polymer experiences a greater range of configurations, since $\langle r_G^2 \rangle \propto \Lambda(T)^2$, where $\Lambda(T)$ is the de Broglie thermal wavelength in Eq~(\ref{eq:debroglie1}) \cite{liqph}. This `swelling' is directly analogous to quantum effects becoming more pronounced at low temperatures, with a greater uncertainty in the position of the particle and hence a greater variety of locations on the potential surface that the ring polymer could be experiencing. 

\section{Ring Polymer Reaction Rate Theory}
\label{sec:rpmd}
RPMD rate theory takes the classical isomorphism and applies it to the quantum flux-side correlation function in Eq~(\ref{eq:cfs_quantum}), producing a rate theory which incorporates physically desirable features of the exact quantum calculation (for example, by including some quantum mechanical effects and independence from the dividing surface) with the comparative computational ease of a classical dynamics calculation, albeit in an extended phase space. The RPMD rate is defined as \cite{Mano_cen}
\begin{equation}
k^{RPMD}(T) = \frac{1}{Q_r(T)} \lim_{t \rightarrow \infty} c_{fs}(t),
\label{eq:kRPMD}
\end{equation}
where the flux-side correlation function is now expressed using the classical isomorphism as \cite{Mano_cen}
\begin{equation}
c_{fs}(t) = \frac{1}{\left(2 \pi \hbar\right)^n}\int d{\bf p_0} \int d{\bf q_0} \ e^{-\beta_n H_n({\bf p_0,q_0})} \delta(\bar q_0 - q^{\ddag})\frac{\bar p_0}{m}h(\bar q_t-q^{\ddag}).
\label{cfs_rpmd}
\end{equation}
This is very similar to the classical rate constant in Eq (\ref{cfs_clas}), except the integrals are over ring-polymer phase space and the positions and momenta in the integrand are those of the centroid,
\begin{eqnarray}
\bar q & = & \frac{1}{n}\sum_{j=1}^n q_j, \\
\bar p & = & \frac{1}{n}\sum_{j=1}^n p_j.
\end{eqnarray}
By taking the infinite-time limit recrossing is allowed to occur, and is captured by the $h(\bar q_t-q^{\ddag})$ factor. The presence of this factor implies that the calculation involves both sampling an initial configuration of the system $({\bf p_0, q_0})$, and evolving it by classical molecular dynamics. 

The resulting rate, $k^{RPMD}(T)$, can be shown to be rigorously independent of the choice of the dividing surface, $\bar q = q^{\ddag}$, exact for a parabolic barrier and in the high temperature limit, equal to the well-known centroid density Quantum Transition State Theory (QTST) rate (detailed below) in the $t \rightarrow 0_+$ limit, and bounded above by this QTST rate \cite{Mano_cen, Mano_pin}. The independence from the choice of dividing surface is particularly important in complex, multidimensional systems where location of the optimum dividing surface is typically difficult to determine, and it is a feature absent from other approximations to quantum mechanical rate constants, such as the Wigner model \cite{wigner} or the quantum instanton (QI) model \cite{qi}. 

The ostensibly arbitrary choice of setting the bead mass equal to the particle mass has been shown to produce greater accuracy in the short time limit \cite{braams} than centroid molecular dynamics \cite{cmd} which takes a similar approach, but has a different choice of bead masses. Richardson and Althorpe have also shown that this choice of fictitious bead masses is essential for the accuracy of the resulting rate coefficient in the deep quantum tunnelling regime \cite{althorpe}. 

Unlike the exact quantum rate in section~\ref{sec:QMratetheory}, where phase information in the $e^{\pm i\hat Ht/\hbar}$ operators is incorporated, this is not included in the classical-like ring polymer rate calculation. Consequently, RPMD rate theory is unlikely to reproduce exact quantum results in models where there is coherent transition state recrossing dynamics, such as a system-bath model for proton transfer at very low temperatures \cite{makrip}. 

\section{Quantum Transition State Theory}
\label{sec:qtst}
I noted earlier that the exact quantum mechanical flux-side correlation function tends to zero as $t \rightarrow 0_+$, thereby precluding a rigorous version of QTST. However, the classical isomorphism of a ring polymer leads naturally to a centroid density QTST rate \cite{miller_qtst},
\begin{eqnarray}
k^{QTST}(T) & = & \frac{1}{Q_r(T)} \lim_{t \rightarrow 0_+} c_{fs}(t) \nonumber \\
& = & \frac{1}{Q_r(T)}\frac{1}{\left(2 \pi \hbar\right)^n}\int d{\bf p} \int d{\bf q} \ e^{-\beta_n H_n({\bf p,q})} \delta(\bar q - q^{\ddag})\frac{\bar p}{m}h\left(\bar p\right).
\label{eq:kqtst}
\end{eqnarray}
By similar arguments to those used for classical transition state theory in Eqs~(\ref{eq:k_clas}--\ref{clas_tst}), this can be shown to be equal to the classical flux through the dividing surface multiplied by the ratio of quantum mechanical dividing surface and reactant partition functions \cite{Mano_cen},
\begin{equation}
k^{QTST}(T) = \frac{1}{2} \langle | \dot q | \rangle \frac{Q_{\ddag}(T)}{Q_r(T)},
\end{equation}
which is precisely the Voth-Chandler-Miller formulation of QTST \cite{miller_qtst}. By taking ring polymer phase space literally, RPMD rate theory is to this QTST what classical rate theory is to classical TST. Here $Q_{\ddag}(T)$ corresponds to the partition function of a ring polymer whose centroid is constrained to the dividing surface,
\begin{equation}
Q_{\ddag}(T) = \frac{1}{\left(2 \pi \hbar\right)^n}\int d{\bf p}  \int d{\bf q} \ e^{-\beta_n H_n({\bf p,q})} \delta(\bar q - q^{\ddag}),
\label{eq:qdag}
\end{equation} 
and $Q_r(T)$, defined in Eq~(\ref{eq:debroglie1}), is the (quantum mechanical and classical) partition function per unit length in the reaction region. 
Like classical transition state theory, the QTST rate can be evaluated in a classical phase space (albeit in an extended dimensionality), but it also includes some quantum mechanical tunnelling effects by nature of the classical isomorphism. Unfortunately QTST is exponentially sensitive to the position of the dividing surface for the same reasons as for classical transition state theory, as given in Eq~(\ref{eq:exp_sen}). Because the QTST rate is always greater than or equal to the full RPMD rate, one can variationally optimise $q^{\ddag}$ by choosing the value which produces the lowest $k^{QTST}(T)$ \cite{varopt}. However, variational optimisation of $q^{\ddag}$ is no guarantee that QTST will give the correct rate if recrossing effects are important \cite{CH_4}.

\section{Previous applications of RPMD rate theory}
The first application of RPMD (before RPMD rate theory was invented) was to the position autocorrelation function of a one-dimensional potential \cite{Mano1}, where good results were obtained if the potential was relatively harmonic. It was also applied to quantum diffusion in liquid para-hydrogen and water, as well as inelastic neutron scattering \cite{liqph, liqh2o, neu_scat}. 

RPMD rate theory was initially applied to the one dimensional Eckart Barrier where $V(q) = V_0{\rm sech}^2(q/a)$ and its asymmetric analogue \cite{Mano_cen}. The centroid version of QTST (section \ref{sec:qtst}) gave answers close to the exact RPMD limit for the optimized dividing surface, as would be expected for a one-dimensional barrier transmission model with a well-defined dividing surface and as such, minimal recrossing. There was also an application to a standard system-bath model for proton transfer in solution \cite{Mano_pin}, which involved a quartic double well potential bilinearly coupled to a harmonic bath with the Ohmic spectral density.

Since then, RPMD rate theory has been applied to more complex reactions, such as a more sophisticated model for acid-base proton transfer in a polar solvent \cite{proton_transfer}, and the quantum diffusion of hydrogen and muonium atoms in liquid water and hexagonal ice \cite{h_mu_diffusion}. This last case was particularly successful, even at a temperature of 8 K and for very light muonium, correctly capturing quantum mechanical (deep tunnelling) effects in the experimental rate coefficient for intercavity hopping.

RPMD rate theory has also been applied to several gas phase reactions, including H + H$_2$, Cl + HCl and F + H$_2$ \cite{bi_rate}. This application compared RPMD to exact quantum results. Even in deep tunnelling (200 K), RPMD rate theory was found to give the correct rates to within a factor of at most three, whereas the corresponding classical calculations underestimated the rates by a factor greater than one thousand. More recently, RPMD has been used to calculate the rate for the more complex reaction H + CH$_4 \rightarrow$ H$_2$ + CH$_3$, producing better results than any other approximate quantum mechanical method \cite{CH_4}.

An interesting facet of the above investigations is that --- in deep tunnelling --- RPMD rate theory has invariably been found to slightly underestimate the rates of symmetric reactions (e.g.\ the H + H$_2$ reaction or symmetric Eckart barrier) and overestimate the rates of asymmetric reactions (e.g.\ H + CH$_4$ or the asymmetric Eckart Barrier). Richardson and Althorpe have recently linked RPMD rate theory with semi-classical instanton theory \cite{semiclas_in}, and in doing so have explained this phenomenon, and justified why RPMD works so well, even at very low temperatures \cite{althorpe}.
\chapter{Non-adiabatic generalization}
\label{chap:nag}
Having summarized previous applications of RPMD theory, all of which involve the ring polymer moving over a single potential energy surface, the theory is now extended to electronically non-adiabatic systems. The only modification required to the theory, as we shall see, is in the form of the ring-polymer potential. Brief calculational details are given, followed by considering how the generalized theory behaves in the high temperature limit, for a single electronic surface, and for a potential which is partly independent of electronic state occupancy. For convenience, one-dimensional notation is still used.
\section{The effective potential}
\label{sec:effpot}
The Hamiltonian, which here contains more than one electronic state, enters RPMD rate theory through the Boltzmann factor $e^{-\beta \hat H}$. For a one-dimensional electronically coupled problem, the potential energy operator, $\hat V(q)$, must include contributions from all electronic states, denoted $|j\rangle$,
\begin{equation}
\hat V(q) = \sum_{j,j'} |j\rangle V_{jj'}(q) \langle j'|.
\label{eq:Vq}
\end{equation}
Here we explore the effect of the non-adiabatic potential in Eq~(\ref{eq:Vq}) on the partition function $Z_n$, which leads to a new form of the ring-polymer Hamiltonian. This is applied to the flux-side correlation function in Eq~(\ref{cfs_rpmd}) to produce an RPMD rate formula which includes electronically non-adiabatic effects. 

Using the form of the partition function in Eq~(\ref{eq:symsplit}), and evaluating the trace in the basis of position and electronic states,
\begin{equation}
 Z_n = \int dq_1 \sum_{j_1} \langle q_1 j_1 | \left(e^{-\beta_n \hat V} \ e^{-\beta_n \hat T} \right)^n | q_1 j_1 \rangle.
\end{equation}
Inserting unit operators $\int q_i \sum_{j_i} | q_i j_i \rangle \langle q_i j_i | $, $i=2,3,\ldots, n$, between each $e^{-\beta_n \hat V} \ e^{-\beta_n \hat T}$ term,
\begin{equation}
 Z_n = \int dq_1 \ldots \int dq_n \sum_{j_1} \ldots \sum_{j_n} 
\langle q_1 j_1 | e^{-\beta_n \hat V} \ e^{-\beta_n \hat T} | q_2 j_2 \rangle \ldots 
\langle q_n j_n | e^{-\beta_n \hat V} \ e^{-\beta_n \hat T} | q_1 j_1 \rangle.
\label{eq:zvt}
\end{equation}
Considering one of these $n$ terms, and noting $\hat T$ is independent of the electronic state,
\begin{equation}
 e^{-\beta_n \hat T} | j \rangle = | j \rangle e^{-\beta_n \hat T},
\end{equation}
and that $\hat V$ is diagonal in the co-ordinate representation (though not in the electronic space),
\begin{equation}
 \langle q | e^{-\beta_n \hat V} = e^{-\beta_n \hat V(q)} \langle q |,
\end{equation}
the bra-kets in Eq~(\ref{eq:zvt}) can be rewritten,
\begin{equation}
\langle q\ j | e^{-\beta_n \hat V} \ e^{-\beta_n \hat T} | q' j' \rangle = 
\langle j | e^{-\beta_n \hat V(q)} | j' \rangle \langle q | e^{-\beta_n \hat T} | q' \rangle.
\end{equation}
Consequently, the partition function will be the product of $\langle j | e^{-\beta_n \hat V(q)}  |j' \rangle$ and $\langle q |  e^{-\beta_n \hat T} |q' \rangle$ terms. The terms involving the momentum operator can be evaluated as in Eqs~(\ref{eq:pqint}--\ref{eq:partition_function}) producing
\begin{eqnarray}
 & Z_n = & \frac{1}{(2\pi\hbar)^n} \int d{\bf p} \int d{\bf q} \ e^{-\beta_n\sum_{j=1}^{n}[p_j^2/2m + \frac{1}{2}m\omega_n^2(q_j-q_{j+1})^2]} \nonumber \\
&\ & \times \sum_{j_1} \ldots \sum_{j_n} \langle j_1 | e^{-\beta_n \hat V(q_1)}| j_2 \rangle \ldots \langle j_n |e^{-\beta_n \hat V(q_n)}| j_1 \rangle.
\label{eq:Zn}
\end{eqnarray}
In order to incorporate the non-adiabatic potential into the Hamiltonian, the multiple summation in Eq~(\ref{eq:Zn}) is trivially rewritten as the exponential of its logarithm, such that
\begin{equation}
 Z_n = \frac{1}{(2\pi\hbar)^n} \int d{\bf p} \int d{\bf q}\ e^{-\beta_n H_n({\bf p, q})},
\end{equation}
where
\begin{equation}
 H_n({\bf p,q}) = \sum_{j=1}^{n} \left[\frac{p_j^2}{2m} + \frac{1}{2}m\omega_n^2(q_j-q_{j+1})^2\right] + V_n({\bf q}),
\label{eq:naham}
\end{equation}
and
\begin{equation}
 V_n({\bf q}) = - \frac{1}{\beta_n} \log  \left[\sum_{j_1} \ldots \sum_{j_n} \langle j_1 | e^{-\beta_n \hat V(q_1)}| j_2 \rangle \ldots \langle j_n |e^{-\beta_n \hat V(q_n)}| j_1 \rangle \right].
\label{eq:napotential}
\end{equation}
Utilising the Hamiltonian in Eq~(\ref{eq:naham}), the reaction rate of a quantum particle in a system with multiple electronic states can be calculated using the flux-side correlation method outlined in section \ref{sec:rpmd}. Usefully, the calculation proceeds in $n$-dimensional classical phase space, as for reactions on a single electronic surface. 
A similar expression to Eq~(\ref{eq:napotential}) has been derived by Schwieters and Voth \cite{schwieters}, though it was used to calculate a path-integral QTST rate and not for evaluation of real-time dynamics nor RPMD reaction rates.%

Unlike the potential for a single electronic surface in chapter~\ref{chap:rpmd}, the force on one bead, $-dV_n({\bf q})/dq_j$, is a function of the positions of all beads, presenting a much greater computational challenge. The form of the potential also prevents a simple graphical picture of a non-adiabatic potential surface. Despite the algebraic complexity, the order of approximation in deriving Eq~(\ref{eq:napotential}) is no greater than that used in the adiabatic case, and arises solely from the symmetric splitting of the Boltzmann operator in Eq~(\ref{split}) leading to a global error of $O(\beta_n^2)$. 

\section{Implementation}%
\label{sec:imp}
Here I consider the computation of the potential in Eq~(\ref{eq:napotential}) and the calculation of the force on each bead, which is required for evolution of the ring polymer in the extended classical phase space.

The electronic potential in Eq~(\ref{eq:Vq}) can be written as a matrix,
\begin{equation}
\langle j | \hat V(q) | j' \rangle =  \left[ {\bf V}(q)\right ]_{jj'}.
\label{eq:vmat}
\end{equation}
The eigenvalues of the matrix ${\bf V}(q)$ are the energies of the electronic eigenstates for a particular value of $q$. The bra-kets in Eq~(\ref{eq:napotential}) therefore correspond to elements of exponential matrices, such that for bead $k$ with position $q_k$,
\begin{equation}
 \langle j | e^{-\beta_n \hat V(q_k)} | j' \rangle = \left(e^{-\beta_n {\bf V}(q_k)}\right)_{jj'} = \left( {\bf M}_k (q_k) \right)_{jj'},
\label{Exp_Matrix}
\end{equation}
where ${\bf M}_k (q_k) \equiv e^{-\beta_n {\bf V}(q_k)}$. 
The trace of the product of these exponential matrices can then be used to evaluate the potential,
\begin{equation}
 V_n({\bf q}) = - \frac{1}{\beta_n} \log\ {\rm Tr} \left[ {\bf M}_1(q_1){\bf M}_2(q_2) \ldots {\bf M}_n(q_n) \right].
\label{pot_mat}
\end{equation}
In order to evolve the ring polymer along a trajectory through the extended classical phase space, the force on each bead, namely the negative derivative of the potential, is required:
\begin{equation}
 -\frac{d}{dq_k} V_n({\bf q}) = +\frac{1}{\beta_n} 
\frac { {\rm tr} \left[ {\bf M}_1(q_1){\bf M}_2(q_2) \ldots {\bf M}_{k-1}(q_{k-1}){\bf D}_k(q_k){\bf M}_{k+1}(q_{k+1}) \ldots {\bf M}_n(q_n) \right]} 
{{\rm tr} \left[ {\bf M}_1(q_1){\bf M}_2(q_2) \ldots {\bf M}_n(q_n) \right]},
\label{eq:force}
\end{equation}
where
\begin{equation}
 {\bf D}_k(q_k) = \frac{d}{dq_k} {\bf M}_k(q_k).
\label{deriv}
\end{equation}
The computation of an exponential matrix and its derivative is outlined in Appendix~\ref{a:expm}.
Noting the cyclic permutation permissible within a trace, the numerator of Eq~(\ref{eq:force}) can be rewritten,
\begin{eqnarray}
 & \ & {\rm tr} \left[ {\bf M}_1(q_1){\bf M}_2(q_2) \ldots {\bf M}_{k-1}(q_{k-1}){\bf D}_k(q_k){\bf M}_{k+1}(q_{k+1}) \ldots {\bf M}_n(q_n) \right] \nonumber \\
 & \ & \qquad = {\rm tr} \left[{\bf D}_k (q_k) {\bf H}_k (q_k)\right],
\end{eqnarray}
where $ {\bf H}_k(q_k)$ is a `hole' matrix,
\begin{equation}
 {\bf H}_k(q_k) = {\bf M}_{k+1}(q_{k+1}) \ldots {\bf M}_n(q_n){\bf M}_1(q_1) \ldots {\bf M}_{k-1}(q_{k-1}).
\label{eq:hole}
\end{equation}
The exponential matrices $ \{ {\bf M}_{k}(q_{k}) \}$ do not commute, so a hole matrix must be evaluated for each bead. From Eq~(\ref{eq:hole}) this would appear to take $n(n-2)$ matrix multiplications, but using Bell's algorithm\footnote{This algorithm, attributed to Martin Bell, is outlined in Appendix~\ref{a:bell}.} only $3n - 6$ multiplications are required.
\section{Discussion}
Despite the complexity of the potential in Eq~(\ref{eq:napotential}), a qualitative physical interpretation is possible by considering various limits in which the dependence of the potential on the electronic eigenstates can be presented analytically. We shall also investigate the case of a single electronic potential energy surface, and a `mixed' potential, only part of which is dependent on the electronic state.

\subsection{The one-bead limit}
\label{sec:1blimit}
In the high temperature limit where the system can be accurately represented by a single bead, $n=1$ and position is expressed as the scalar $q$ rather than the ring polymer vector ${\bf q}$. The non-adiabatic potential in Eq~(\ref{eq:napotential}) reduces to
\begin{equation}
V_1(q) = -\frac{1}{\beta} \log \sum_{j} \langle j | e^{-\beta \hat V(q)}| j \rangle.
\end{equation}
Considering a two-level system and evaluating the trace in the basis of the eigenstates of ${\bf V}(q)$, with energies $E_1(q)$ and $E_2(q)$,
\begin{equation}
V_1(q) = -\frac{1}{\beta} \log \left( e^{-\beta E_1(q)} + e^{-\beta E_2(q)} \right).
\label{eq:v1}
\end{equation}
Considering the force,
\begin{eqnarray}
F(q) = -\frac{dV_1(q)}{dq} & = & +\frac{1}{\beta} \frac{-\beta E_1'(q)e^{-\beta E_1(q)} + -\beta E_2'(q) e^{-\beta E_2(q)}}{e^{-\beta E_1(q)} + e^{-\beta E_2(q)}}  \\
& = & -\frac{E_1'(q)e^{-\beta E_1(q)} + E_2'(q) e^{-\beta E_2(q)}}{e^{-\beta E_1(q)} + e^{-\beta E_2(q)}} \\
& = & F_1(q)P_1(q) + F_2(q)P_2(q),
\label{eq:1bforce}
\end{eqnarray}
where $F_i(q) = -E_i'(q)$ is the force on a bead solely experiencing electronic potential surface $i$, and 
\begin{equation}
 P_i(q) = \frac{e^{-\beta E_i(q)}}{e^{-\beta E_1(q)} + e^{-\beta E_2(q)}},
\end{equation}
is the thermal probability of the bead being found on surface $i$. The overall force in Eq~(\ref{eq:1bforce}) is therefore a Boltzmann-weighted average of the forces from the two electronic potential energy surfaces, as would be physically expected.

Equation (\ref{eq:v1}) can be rewritten as
\begin{eqnarray}
V_1(q) & = &  -\frac{1}{\beta} \log \left[ e^{-\beta E_1(q)}\left(1 + e^{-\beta (E_2(q)-E_1(q))}\right) \right]  \\
& = & E_1(q) -\frac{1}{\beta} \log \left(1 + e^{-\beta (E_2(q)-E_1(q))}\right)  \\
& = & E_1(q) -\frac{1}{\beta} \log \left(1 + e^{-\beta \Delta E(q)}\right),
\end{eqnarray} 
where $\Delta E(q) = E_2(q) - E_1(q)$. 
Similarly,
\begin{equation}
 P_1(q) = \frac{1}{1 + e^{-\beta \Delta E(q)}}, \ P_2(q) = \frac{e^{- \beta \Delta E(q)}}{1 + e^{-\beta \Delta E(q)}},
\end{equation}
and
\begin{equation}
 F(q) = \frac{F_1(q) + e^{-\beta \Delta E(q)}F_2(q)}{1 + e^{-\beta \Delta E(q)}}.
\end{equation}
The advantage of the variable $\Delta E(q)$ is that it can be set to zero in the limit of degenerate states, and taken to infinity in the limit of widely-separated states, which we now explore.%
\subsubsection{Degenerate states}%
If, for some value of $q$, the electronic eigenstates of ${\bf V}(q)$ are of equal energy, \mbox{$\Delta E(q) = 0$}. In this limit,
\begin{eqnarray}
V_1(q) & = & E_1(q) -\frac{1}{\beta} \log 2, \label{eq:ds1} \\
P_1(q) & = & P_2(q) \ = \ \frac{1}{2}, \label{eq:ds2} \\
F(q) & = & \tfrac{1}{2}\left(F_1(q) + F_2(q)\right). \label{eq:ds3}
\end{eqnarray} 
The $\log 2$ term in Eq~(\ref{eq:ds1}) can be understood as the entropy arising from the occupation of two degenerate states, which are equally occupied by Eq~(\ref{eq:ds2}), such that the force on the bead is the average of the forces from each electronic potential surface, as given by Eq~(\ref{eq:ds3}).

\subsubsection{Widely separated states}
Conversely, when state $i=2$ is far higher in energy than state $i=1$, $\beta \Delta E(q) \rightarrow \infty$ and
\begin{eqnarray}
V_1(q) & = & E_1(q) -\lim_{\Delta E(q) \rightarrow \infty} \frac{1}{\beta} \log \left(1 + e^{-\beta \Delta E(q)}\right) \nonumber \\
& \simeq & E_1(q),
\label{eq:bigE} \\
P_1(q) & \rightarrow & 1,  \\
F(q) & \rightarrow & F_1(q).
\end{eqnarray}
In this regime, the physics of the system is dominated by the properties of the lowest electronic state. This is a physically plausible, desirable feature, as the statistical probability of occupying a high-lying state is very low. As the states approach each other, the mixing increases. 
\subsection{Single electronic state}
The derivations in section \ref{sec:effpot} were completely general in terms of the number of electronic states present in the system. For a single electronic state, Eq (\ref{eq:Vq}) simplifies to
\begin{equation}
\hat V(q) = |j\rangle V_{jj}(q) \langle j|.
\end{equation}
The sums over $j$ and $j'$ disappear as there is only one electronic state, making the potential a scalar quantity. The non-adiabatic potential in Eq~(\ref{eq:napotential}) reduces to
\begin{eqnarray}
 V_n({\bf q}) & = & - \frac{1}{\beta_n} \log  \left[ e^{-\beta_n V(q_1)} \ldots e^{-\beta_n V(q_n)}\right] \\
& = & - \frac{1}{\beta_n} \log  \left[ e^{-\beta_n \sum_{k=1}^{n} V(q_k)}\right] \\
& = & \sum_{k=1}^{n} V(q_k),
\end{eqnarray}
which agrees exactly with the adiabatic ring polymer potential in Eq~(\ref{eq:adv}), as would be expected. As all matrices associated with the potential become scalars, the force calculation in Eq~(\ref{eq:force}) simplifies too,
 \begin{eqnarray}
-\frac{d}{dq_k} V_n({\bf q}) & = & +\frac{1}{\beta_n} 
\frac { M_1(q_1)\ldots M_{k-1}(q_{k-1})D_k(q_k)M_{k+1}(q_{k+1}) \ldots M_n(q_n) } 
{M_1(q_1)M_2(q_2) \ldots M_n(q_n)} \\
& = & +\frac{1}{\beta_n}\frac{D_k(q_k)}{M_k(q_k)} \\
& = & +\frac{1}{\beta_n}\frac{\frac{d}{dq_k} e^{-\beta_n V_k(q_k)}}{e^{-\beta_n V_k(q_k)}} \\
& = & -\frac{d}{dq_k} V_k(q_k),
\end{eqnarray}
as would also be expected from the adiabatic potential presented in Eq~(\ref{eq:adv}).

\subsection{Mixed potential}
\label{subsec:mixp}
The potential of some systems, such as the spin-boson model explored in chapter~\ref{chap:sb}, may be expressed as a sum of two terms, one of which, $V^{elec}(q)$, is electronically coupled, and the other independent of electronic state,
\begin{equation}
 \hat V(q) = \sum_{j,j'} |j\rangle V^{elec}_{jj'}(q) \langle j'| + V^{0}(q),
\label{eq:vmix}
\end{equation}
where $V^{0}(q)$ is the electronically diagonal component of the potential, such that 
\begin{equation}
 \langle j'| V^{0}(q) = V^{0}(q) \langle j'|.
\end{equation}
The bra-kets in the non-adiabatic potential Eq~(\ref{eq:napotential}) can be rewritten,
\begin{equation}
 \langle j | e^{-\beta_n \hat V(q)}| j' \rangle = \langle j | e^{-\beta_n \hat V^{elec}(q)}| j' \rangle \times e^{-\beta_n V^{0}(q)},
\end{equation}
such that the non-adiabatic potential becomes
\begin{eqnarray}
 V_n({\bf q}) &=& - \frac{1}{\beta_n} \log \left[\sum_{j_1}\ldots\sum_{j_n} \langle j_1 | e^{-\beta_n \hat V^{elec}(q_1)}| j_2 \rangle\ldots\langle j_n |e^{-\beta_n \hat V^{elec}(q_n)}| j_1 \rangle \right. \nonumber \\
 &\  & \left. \times \prod_{j=1}^{n} e^{-\beta_n V^{0}(q_j)} \right] \\
& = & -\frac{1}{\beta_n} \log  \left[\sum_{j_1}\ldots\sum_{j_n} \langle j_1 | e^{-\beta_n \hat V^{elec}(q_1)}| j_2 \rangle\ldots\langle j_n |e^{-\beta_n \hat V^{elec}(q_n)}| j_1 \rangle \right] \nonumber\\
& \ & + \sum_{j=1}^{n} V^{0}(q_j)\\
& = & V_n^{elec}({\bf q}) + V_n^{0}({\bf q}).
\end{eqnarray}
Here $V_n^{elec}({\bf q})$ is the electronic component of the potential in Eq~(\ref{eq:vmix}) treated using the electronically non-adiabatic ring polymer potential in Eq~(\ref{eq:napotential}),
\begin{equation}
V_n^{elec}({\bf q}) = - \frac{1}{\beta_n} \log \left[\sum_{j_1} \ldots \sum_{j_n} \langle j_1 | e^{-\beta_n \hat V^{elec}(q_1)}| j_2 \rangle \ldots \langle j_n |e^{-\beta_n \hat V^{elec}(q_n)}| j_1 \rangle \right],
\end{equation}
and $V_n^{0}({\bf q})$ is the remaining component of the potential in Eq~(\ref{eq:vmix}) treated using the adiabatic ring polymer potential in Eq~(\ref{eq:adv}),
\begin{equation}
 V_n^{0}({\bf q}) = \sum_{j=1}^n V^{0}(q_j).
\end{equation}
This means that in a complicated problem, the electronically non-adiabatic part of the potential can be calculated separately to any part of the potential which does not depend on the electronic states, and likewise for the forces on the ring polymer beads. The advantage of this approach lies in computational efficiency, since the evaluation of the simple sum for adiabatic contributions to the potential is computationally trivial compared to the difficulties in calculating the electronically non-adiabatic potential explained in section~\ref{sec:imp} and appendices~\ref{a:expm} and \ref{a:bell}.

\section{Summary}
In this chapter I have generalized RPMD rate theory to electronically non-adiabatic systems. It was then shown that the ostensibly complex formulation of the non-adiabatic ring polymer potential in Eq~(\ref{eq:napotential}) produces physically plausible answers in the one-bead, high temperature limit and reduces to the conventional adiabatic formulation of the potential in the limit of only one electronic state. Furthermore, contributions to the potential which do not depend on electronic state can be considered separately to those which do. 
\clearpage
\chapter{One-Dimensional Model}
\label{chap:1D}
Having examined conventional RPMD rate theory in chapter \ref{chap:rpmd}, and extended it to a system with multiple electronic surfaces in chapter \ref{chap:nag}, the rate theory is now applied to two one-dimensional barrier transmission models. These models are chosen due to their non-adiabatic electronic effects and nuclear tunnelling, providing an initial test of the generalization of RPMD to systems with more than one electronic surface. The idea of a non-adiabatic transition between two electronic states, whose energy changes with reaction co-ordinate, has existed since at least the 1930s \cite{Zener}, named the Landau-Zener model after its early proponents. 

A symmetric model is initially considered, followed by an asymmetric one where the location of the optimum dividing surface cannot be inferred by symmetry. As mentioned earlier, asymmetric and symmetric barrier transmission systems are qualitatively different for RPMD rate theory; previous applications to single surface reactions have observed an underestimation of the exact rate for symmetric potentials and overestimation of the rate for asymmetric potentials \cite{CH_4, bi_rate}, as later explained by Richardson and Althorpe \cite{althorpe}.

\section{Symmetric curve-crossing}

\subsection{The symmetric model}
Using the matrix form of the non-adiabatic potential in Eq~(\ref{eq:vmat}), the first model I shall consider is defined by
\begin{equation}
{\bf V}(q) = 
\left( \begin{array}{cc}
        Ae^{+Bq} & C        \\
	C        & Ae^{-Bq} \\
       \end{array}\right),
\end{equation}
where $A = 0.02$, $B = 2.0$ and $C = 0.004$, all in atomic units. The particle mass is 2000 atomic units, close to that of a hydrogen atom (1837~a.u.), ensuring tunnelling effects are present at ambient temperatures and below. The diabatic and adiabatic curves are shown in Fig~\ref{symcurve}.

\newlength{\figwidth}
\setlength{\figwidth}{\columnwidth}

\begin{figure}[h]
\centering
\resizebox{\figwidth}{!} {\includegraphics[angle = 270]{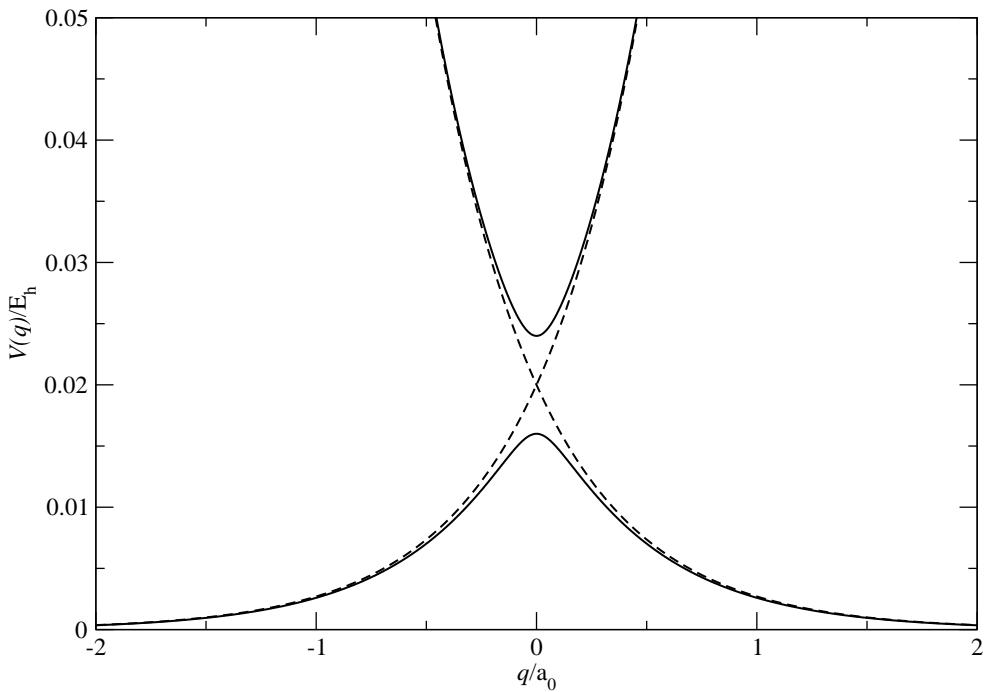}}
\caption{Diabatic (dashed lines) and adiabatic (solid lines) curves for the symmetric curve-crossing model.}
\label{symcurve}
\end{figure}

Although the RPMD rate constant is independent of dividing surface location, the convergence of the flux-side correlation function is greatly enhanced with an appropriate choice of the dividing surface \cite{CH_4}. By symmetry, the optimum dividing surface (where the centroid potential of mean force is maximised) is at $q^{\ddag} = 0$. 

Exact quantum mechanical transmission probabilities for this symmetric model are calculable by the log derivative method \cite{log_deriv, early_log_deriv}, and are illustrated in Fig~(\ref{nesym}).

\begin{figure}[h]
\centering
\resizebox{\figwidth}{!} {\includegraphics[angle = 270]{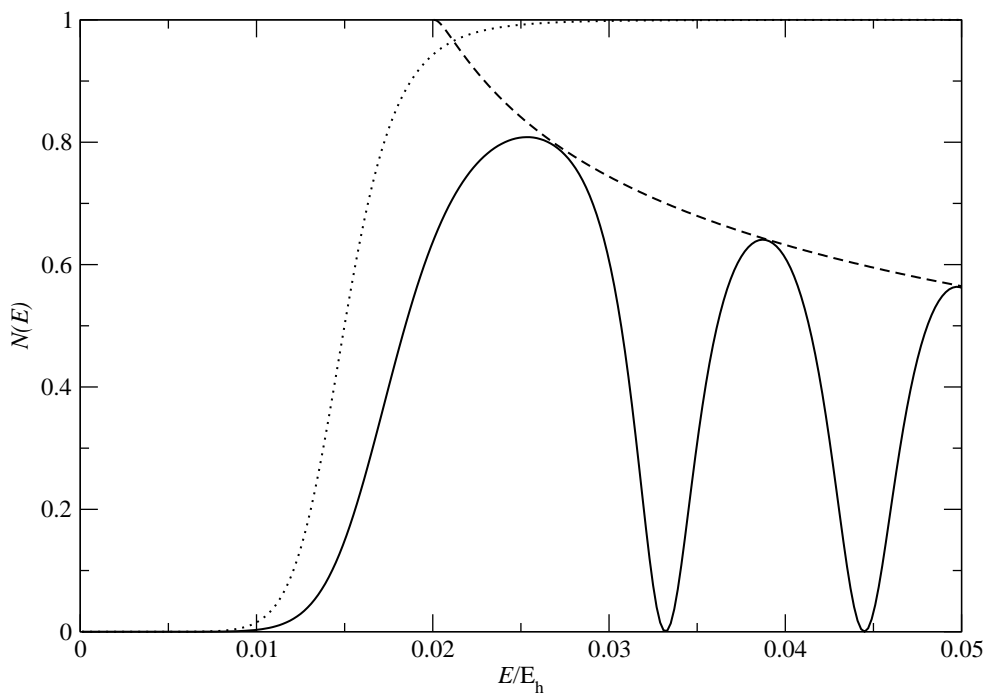}}
\caption{Microcanonical reaction probabilities $N(E)$ for the symmetric model as a function of microcanonical energy $E$. The dotted line ($\cdots$) illustrates the reaction probability on the ground adiabatic surface, the dashed line (- - -) the envelope of the reaction probability arising from Landau-Zener analysis \cite{Child}, and the solid line (------) the reaction probability arising from reaction on the coupled, non-adiabatic surfaces. The programs for producing these results were provided by D. E. Manolopoulos.}
\label{nesym}
\end{figure}

The adiabatic reaction probability increases from zero to unity as $E$ becomes greater than the barrier height (0.016 a.u.). By contrast, the transmission probability for the non-adiabatic case shows Landau-Zener-St\"uckelberger oscillations \cite{Child} arising from quantum mechanical interference between competing pathways. 

The exact quantum mechanical thermal rate constant can be calculated by taking a Boltzmann average over microcanonical reaction probabilities \cite{Mano_notes},
\begin{equation}
 k(T) = \frac{1}{Q_r(T)} \ootph \int_0^{\infty} dE \ e^{-\beta E}N(E),
\label{eq:intne}
\end{equation}
where $Q_r(T)$ is defined in Eq~(\ref{eq:debroglie1}). One can qualitatively infer from Eq~(\ref{eq:intne}) and Fig~\ref{neasym} that the thermal non-adiabatic rate constant will be lower than the adiabatic one for all temperatures, due to a lower or equal reaction probability for all values of $E$.

\subsection{Computational Details}
\label{subsec:comp}
As outlined in Eqs~(\ref{eq:kRPMD}) and (\ref{cfs_rpmd}), the RPMD rate is
\begin{equation}
k^{RPMD}(T) = \frac{1}{Q_r(T)} \lim_{t \rightarrow \infty} \frac{1}{\left(2 \pi \hbar\right)^n}\int d{\bf p_0} \int d{\bf q_0} \ e^{-\beta_n H_n({\bf p_0,q_0})} \delta(\bar q_0 - q^{\ddag})\frac{\bar p_0}{m}h\left(\bar q_t-q^{\ddag}\right).
\label{eq:longrate}
\end{equation}
As the Landau-Zener model is a one-dimensional barrier transmission model, the partition function $Q_r(T)$ can be evaluated analytically as the inverse de Broglie thermal wavelength, given in Eq~(\ref{eq:debroglie1}). 

From Eq~(\ref{eq:naham}), the Hamiltonian is 
\begin{equation}
 H_n({\bf p,q}) = \sum_{j=1}^{n} \left[\frac{p_j^2}{2m} + \frac{1}{2}m\omega_n^2(q_j-q_{j+1})^2\right] + V_n({\bf q}),
\end{equation}
which can be expressed entirely in vector notation as 
\begin{equation}
 H_n({\bf p,q})= \frac{\bf p^Tp}{2m} + \frac{1}{2}m {\bf q^T A q} + V_n({\bf q}),
\label{eq:hvec}
\end{equation}
where ${\bf A}$ is a matrix of bead spring interactions. For a free ring polymer, where $V_n({\bf q}) = 0$, Eq~(\ref{eq:hvec}) reduces to
\begin{equation}
 H_n^0({\bf p,q})= \frac{\bf p^Tp}{2m} + \frac{1}{2}m{\bf q^T A q}.
\label{eq:h0}
\end{equation}

In subsection~\ref{sssec:nmodes}, ${\bf A}$ is diagonalised leading to the normal modes of a free ring polymer. These are used to construct a time evolution algorithm in subsection~\ref{sssec:1dte}, and an efficient method for Monte Carlo sampling of initial configurations in subsection~\ref{sssec:mc}.

\subsubsection{Normal modes of a free ring polymer}
\label{sssec:nmodes}
The symmetric matrix $\bf A$ is of the form
\begin{equation}
{\bf A} = 
\omega_n^2 
\left(
\begin{array}{ccccc}
2 & -1 & 0 & \cdots & -1  \\
-1 & 2 & -1 & \      & 0  \\
0 & -1 & 2 & \      & 0  \\
\vdots & \ & \ & \ddots & \vdots  \\
-1 & 0 & 0 & \cdots & 2 
\end{array}
\right).
\end{equation}
This is similar to a H\"uckel Hamiltonian matrix for a cyclic polyene \cite{hueckel}. ${\bf A}$ can be diagonalised using an orthonormal matrix $\bf C$, such that
\begin{equation}
 \bf C^TAC = \tilde \Omega{\rm ^2,}
\end{equation}
where $\bf \tilde \Omega$ is a diagonal matrix corresponding to the frequencies of the normal modes of a free ring polymer \cite{michele}. Noting that $\bf CC^T = C^TC = I$, the free ring polymer Hamiltonian from Eq~(\ref{eq:h0}) can be rewritten as
\begin{eqnarray}
 H_n^0({\bf p,q}) & = & \frac{1}{2m} {\bf p^TCC^Tp} + \frac{1}{2}m{\bf q^TCC^TACC^T q}, \\
& = & \frac{1}{2m} {\bf \tilde p^T\tilde p} + \frac{1}{2}m{\bf \tilde q^T \tilde \Omega}^2{\bf \tilde q}, 
\end{eqnarray}
where $\bf \tilde p$ and $\bf \tilde q$ are vectors of the momenta and co-ordinates of a free ring polymer's normal modes, defined as
\begin{equation}
\bf \tilde p = C^T p,\ \tilde q = C^T q.
\end{equation}
Choosing to number the beads with $j=1,2,\ldots, n$, and the normal modes of a ring polymer with $k = 0, 1,\ldots, n-1$, where $n$ is even, the elements of $\bf C$ are 
\begin{equation}
C_{jk} = 
\left\{ \begin{array}{cc}
\sqrt{\frac{1}{n}}, 	& k = 0 \\
\sqrt{\frac{2}{n}}\cos\left(\frac{2\pi jk}{n}\right),	& 1 \leq k \leq \frac{n}{2} -1 \\
\sqrt{\frac{1}{n}}(-1)^j, & 	k = \frac{n}{2} \\
\sqrt{\frac{2}{n}}\sin\left(\frac{2\pi jk}{n}\right), 	& \frac{n}{2} + 1 \leq k \leq n-1 \\
\end{array}
\right..
\end{equation}
This corresponds to a discrete halfcomplex Fourier Transform with appropriate normalisation\footnote{The transformation matrix can also be written as a conventional Discrete Fourier Transform, but as the $k$th and $(n-k)$th eigenvalues are identical, linear combinations of degenerate normal modes can be taken to produce cosine and sine terms, thereby keeping the transformation real-to-real and improving computational efficiency.},  performed efficiently by standard library routines. The corresponding eigenvalues (diagonal elements of ${\bf\tilde \Omega}^2$) of the $k$th normal mode are
\begin{equation}
\tilde \omega_k^2 = 4\omega_n^2 \sin^2\left(\frac{k\pi}{n}\right),
\label{eq:omegak}
\end{equation}
such that the Hamiltonian for the free ring polymer is decoupled into
\begin{equation}
H_n^0({\bf \tilde p, \tilde q}) = \sum_{k = 0}^{n-1} \left[\frac{\tilde p_k^2}{2m} + \frac{1}{2}m\tilde \omega_k^2 \tilde q_k^2\right],
\label{eq:H0}
\end{equation}
so that the normal modes of the free ring polymer can be evolved analytically by simple harmonic motion.

\subsubsection{Time evolution}
\label{sssec:1dte}
Time evolution is by a symplectic numerical integrator, where the overall Hamiltonian $H_n({\bf p,q})$ is split into a free ring polymer part, $H_n^0({\bf p,q})$, for which evolution is analytic (as described in the section~\ref{sssec:nmodes}), and a potential part $V_n({\bf q})$. The general procedure for time evolution through an interval $\delta t$ is as follows \cite{michele}:
\begin{enumerate}
\item
Evolve momenta under their forces for time $\delta t/2$
\begin{equation}
p_j \leftarrow p_j - \frac{\delta t}{2} \frac{dV_n({\bf q})}{dq_j},
\end{equation}
\item Transform to the normal modes of a free ring polymer
\begin{equation}
\bf \tilde p \leftarrow C^Tp, \qquad \bf \tilde q \leftarrow C^Tq,
\end{equation}
\item 
Evolve the normal modes analytically by time $\delta t$
\begin{equation}
\binom{\tilde p_k}{\tilde q_k} \leftarrow
\left(
\begin{array}{cc}
\cos(\tilde \omega_k \delta t) & -m\tilde \omega_k \sin(\tilde \omega_k \delta t) \\
\frac{1}{m \tilde \omega_k}\sin(\tilde \omega_k \delta t) & \cos(\tilde \omega_k \delta t)
\end{array}
\right)
\binom{\tilde p_k}{\tilde q_k},
\end{equation}
\item Transform to the bead representation
\begin{equation}
\bf p \leftarrow C\tilde p, \qquad \bf q \leftarrow C\tilde q,
\end{equation}
\item Evaluate forces, $-dV_n({\bf q})/dq_k$, with bead positions at time $t + \delta t$,
\item Evolve momenta under new forces for time $\delta t/2$,
\begin{equation}
p_j \leftarrow p_j - \frac{\delta t}{2} \frac{dV_n({\bf q})}{dq_j}.
\end{equation}
\end{enumerate}
Stages 1 and 6 are evolution of the momenta under the potential $V_n({\bf q})$. Stages 2 and 4 involve the transformations described in the subsection~\ref{sssec:nmodes}, and stage 3 is exact evolution of a free ring polymer under $H_n^0({\bf p,q})$, derived by analytically solving the equations of motion for a simple harmonic oscillator. This algorithm conserves phase space volume, satisfying Liouville's theorem, such that $d{\bf p_0}d{\bf q_0} = d{\bf p_t}d{\bf q_t}$ \cite{michele}. Consequently, energy drift is not observed and instead the total calculated energy of the system fluctuates around its starting value. 

The time step, $\delta t$, must be sufficiently small to accurately reproduce trajectories and conserve the Hamiltonian. The time evolution scheme can be justified as a series of exact evolutions under approximate Hamiltonians, and possesses global error $O(\delta t^2)$ \cite{michele}. 
For the dynamics of the highest frequency mode of the ring polymer not to be washed out, the time step should be considerably less than the time period of the highest frequency mode of the free ring polymer, i.e.\ $\delta t < 2\pi/\omega_{max}$. From Eq~(\ref{eq:omegak}), 
\begin{equation}
 \omega_{max} = \tilde \omega_{k=n/2} = 2\omega_n,
\end{equation}
so a `safe' time step is given by
\begin{equation}
 \delta t < \frac{2\pi}{\omega_{max}} = \frac{\pi \hbar}{nk_B T}.
\label{eq:deltat}
\end{equation}
However, the highest frequency mode of the ring polymer may not contribute to the dynamics of the system, and in practice a converged time step is found by decreasing $\delta t$ until the total energy during a trajectory is well conserved. 
%
%

\subsubsection{Monte Carlo Sampling}
\label{sssec:mc}
In order to evaluate the time evolved side operator $h(\bar q_t-q^{\ddag})$ the system is evolved by classical molecular dynamics from a starting configuration denoted $({\bf p_0,q_0})$. Using biased Monte Carlo sampling, one aims to find a sampling function which incorporates as much of the original integrand in Eq~(\ref{eq:longrate}) as possible, but whose distribution can be sampled efficiently. 

I used two similar sampling functions,
\begin{eqnarray}
\rho_+ ({\bf p, q}) & = & h(\bar p) e^{-\beta_n \left[{\bf p^Tp}/2m + \frac{1}{2}m\omega_n^2 {\bf q^TAq}\right]} \delta(\bar q - q^{\ddag})\frac{\bar p}{m}  \label{eq:rhoplus}, \\
\rho_- ({\bf p, q}) & = & h(-\bar p) e^{-\beta_n \left[{\bf p^Tp}/2m + \frac{1}{2}m\omega_n^2 {\bf q^TAq}\right]} \delta(\bar q - q^{\ddag})\frac{\bar p}{m}, \label{eq:rhominus}
\end{eqnarray}
which have the analytic integrals,
\begin{equation}
\int d{\bf p} \int d{\bf q} \ \rho_{\pm}({\bf p, q}) = \pm \frac{(2\pi\hbar)^{n-1}}{\beta}.
\end{equation}
The rate equation~(\ref{eq:longrate}) can be separated into considerations of positive and negative initial momentum centroids. Choosing to multiply Eq~(\ref{eq:longrate}) by $\pm (2\pi\hbar)^{n-1}/\beta$ and divide by $\int d{\bf p} \int d{\bf q} \ \rho_{\pm}({\bf p, q})$, and choosing $\rho_{\pm}({\bf p, q})$ to sample the initial co-ordinates and momenta of the system $({\bf p_0, q_0})$, the rate equation becomes
\begin{eqnarray}
k^{RPMD}(T) & = & \lim_{t \rightarrow \infty} \frac{1}{\sqrt{2\pi\beta m}} \left[
\frac{\int d{\bf p_0} \int d{\bf q_0}\ \rho_+({\bf p_0, q_0}) e^{- \beta_n V_n({\bf q_0})}  h\left(\bar q_t-q^{\ddag}\right)}{\int d{\bf p_0} \int d{\bf q_0} \ \rho_+({\bf p_0, q_0})}\right.  \nonumber \\
& \ &  + \left. 
\frac{\int d{\bf p_0} \int d{\bf q_0}\ \rho_-({\bf p_0, q_0}) e^{- \beta_n V_n({\bf q_0})}  h\left(\bar q_t-q^{\ddag}\right)}{\int d{\bf p_0} \int d{\bf q_0} \ \rho_-({\bf p_0, q_0})} \right].
\label{eq:ratesam}
\end{eqnarray}
Initial momenta and positions are sampled in the normal mode representation of $\rho_{\pm}({\bf p,q})$,
\begin{eqnarray}
\rho_{\pm}({\bf \tilde p, \tilde q}) & = & h(\pm \bar p) e^{-\beta_n \left[{\bf \tilde p^T\tilde p}/2m + \frac{1}{2}m\tilde \omega_k^2{\bf \tilde q^T\tilde q}\right]}\delta(\bar q - q^{\ddag})\frac{\bar p}{m}  \\
& = & h\left(\pm \frac{\tilde p_0}{\sqrt{n}}\right)\frac{\tilde p_0}{\sqrt{n}m} e^{-\beta_n \tilde p_0^2/2m} 
\left( \prod_{k=1}^{n-1} e^{-\beta_n \tilde p_k^2/2m} \right) \\
& \ & \times\ 
\delta\left(\frac{\tilde q_0}{\sqrt{n}} - q^{\ddag}\right)
\left(\prod_{k=1}^{n-1} e^{-\frac{1}{2}\beta_n m\tilde \omega_k^2 \tilde q_k^2} \right).
\end{eqnarray}
The zeroth momentum normal mode, $\tilde p_0 = \sqrt{n}\bar p$, is sampled from the Gamma distribution of order 2, in order to capture the $\tilde p_0  e^{-\beta_n \tilde p_0^2/2m}$ distribution. All other momentum normal modes are sampled from Gaussian deviates with zero mean and standard deviation $\sqrt{m/\beta_n}$. The position centroid $\tilde q_0$ is set equal to $\sqrt{n} q^{\ddag}$, and the $k$th co-ordinate normal modes, $k \neq 0$, are sampled from Gaussian deviates with zero mean and standard deviation $1/(\sqrt{m}\tilde \omega_k)$.

These normal mode co-ordinates and momenta, $({\bf \tilde p_0, \tilde q_0})$, are then transformed to the bead representation $({\bf p_0, q_0})$, and the initial potential of the system $V_n({\bf q_0})$ is calculated, along with the force on each bead $ dV_n({\bf q_0})/dq_j$. The system is evolved to long time, $t_{\infty}$, by classical molecular dynamics (in this model, 60 fs is sufficient), and after each time step, $h(\bar q_t-q^{\ddag})$ is evaluated. If the position centroid is on the products side of the dividing surface, $e^{- \beta_n V_n({\bf q_0})}$ (calculated at the start of the trajectory) is added to $c_{fs}(t) / Q_r(T)$, scaled by $1/\sqrt{2\pi\beta m}$. Trajectories are run in pairs of $({\bf p_0, q_0})$ and $({\bf -p_0, q_0})$ initial configurations in order to sample both positive and negative initial momentum centroids. Plots of $c_{fs}(t)/Q_r(T)$ are analysed to check for convergence with respect to time step, number of beads, and number of trajectories. When converged, the long time limit, $c_{fs}(t_{\infty})/Q_r(T)$ is the rate.

An advantage of the sampling function in Eq~(\ref{eq:ratesam}) is that it produces the exact answer analytically in the `classical' one bead limit with the optimum dividing surface.  In this limit, there is no recrossing, such that negative momentum states always end up on the reactant side of the dividing surface (thereby contributing nothing to the rate equation), and positive momentum states are always found on the products side. The second part of Eq~(\ref{eq:ratesam}) vanishes, and the first part is analytically calculable as 
\label{p:1bclass}
\begin{eqnarray}
k^{cl,TST}(T) & = & \frac{1}{\sqrt{2\pi\beta m}} e^{- \beta V(q^{\ddag})} \\
& = & \tfrac{1}{2} \langle | \dot q | \rangle e^{- \beta V(q^{\ddag})},
\end{eqnarray}
which is the classical transition state theory rate introduced in section \ref{sec:cltst} for a one dimensional model and a potential which tends to zero in the reactant limit ($\lim_{q \to -\infty} V(q) = 0 $). The $e^{-\beta V(q^{\ddag})}$ factor means the classical rate will appear linear on an Arrhenius plot, if $V(q^{\ddag})$ is not itself a function of temperature.\footnote{This is generally true on a single electronic potential energy surface, but not for a non-adiabatic potential, as evident from the $\beta$ dependence of Eq~(\ref{eq:napotential}).}

\subsubsection{Convergence}
\label{sssec:conv}
The parameters which required converging for this calculation were the number of beads $n$, the time step $\delta t$, the length of the trajectories $t_{\infty}$ and the number of trajectories $N$. The calculation increases in accuracy as $n \to \infty$, $N \to \infty$, $t_{\infty} \to \infty$ and $\delta t \to 0$. Convergence is obtained by altering the parameters until the rate remains unchanged to within graphical accuracy.

At the lowest temperature (100 K), 512 beads were sufficient to converge the rate constant, and at 200 K, 256 beads were needed. To model this for any temperature in the range $100\ {\rm K} \leq T \leq 1000\ {\rm K}$, the number of beads was calculated as $51200/T$ rounded down to the nearest power of two\footnote{The Fast Fourier Transform algorithms are particularly efficient if the number of beads is an integer power of two.}. Using the same number of beads for all temperatures would not only be more expensive computationally, but by Eq~(\ref{eq:deltat}) would require very small time steps at high temperatures.

To my surprise, a time step of $\delta t = 2\ {\rm fs}$ was sufficient for convergence of the trajectories, even though it is considerably greater than that suggested by Eq~(\ref{eq:deltat}). This implies that the dynamics of the very high frequency modes of the ring polymer do not interact appreciably with the external potential $V_n({\bf q})$, and consequently do not affect the path of the trajectory. This was checked with a test case using a time step of 1 fs at 200 K.

By inspection of the plots of the flux-side correlation function against time, a choice of $t_{\infty} = 60$ fs was sufficient to produce a converged plateau.

For each temperature, the number of trajectories $N$ was iterated to convergence within a 5\% error, corresponding to an error on the logarithmic rate graph of approximately $\pm0.02$. At the highest temperature (1000 K), 4000 trajectories sufficed, but at 100 K, $4\times10^6$ were required.
 
\subsection{Results}
RPMD and exact quantum reaction rates for the symmetric Landau-Zener model are illustrated in Fig~(\ref{symrate}) as a function of temperature. The programs for producing exact quantum results for this graph, and the asymmetric model, were provided by D.\ E.\ Manolopoulos.
\begin{figure}[htb]
\centering
\resizebox{\figwidth}{!} {\includegraphics[angle = 270]{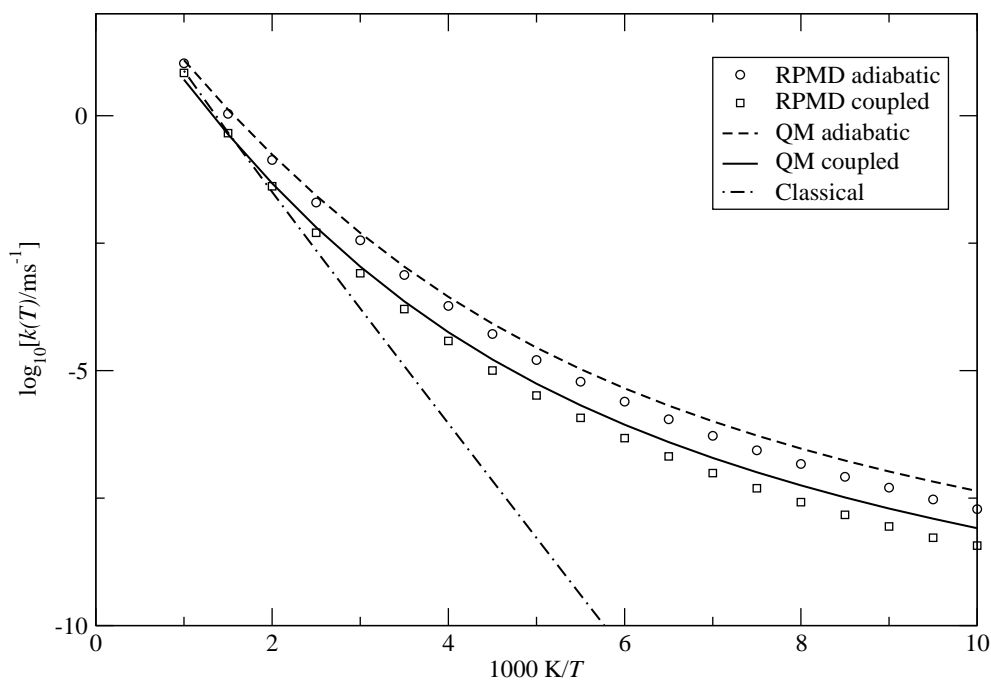}}
\caption{Reaction rates as a function of temperature for the symmetric curve-crossing model.}
\label{symrate}
\end{figure}

The adiabatic rate is also given for comparison, where the ring polymer moves solely on the lower electronic potential energy surface. The `classical' $n=1$ ring polymer bead rate on the coupled surfaces is shown, which is the same to within graphical accuracy as the `classical' rate for a single bead moving only on the lower adiabatic surface surface (not shown). This arises because the temperature of the energy gap between the two electronic states is roughly 2500 K, and as such the upper adiabatic potential surface makes hardly any contribution to the potential a single bead experiences, in accordance with Eq~(\ref{eq:bigE}).

The RPMD rate becomes exactly equal to the quantum rate in the adiabatic case for high $T$, as expected \cite{Mano_pin}. In this limit, the ring polymer collapses into a single bead, for which the rate calculation mirrors that for classical transition state theory with the optimum dividing surface (as detailed in section~\ref{sec:cltst}). As the temperature falls, the classical rate, which has an Arrhenius-like temperature dependence, vastly underestimates the quantum rate but the RPMD rate remains accurate roughly to within a factor of two, even in the deep tunnelling regime at very low temperatures. 
For the reaction on the coupled electronic surfaces, the RPMD rate does not tend to the exact quantum rate for high $T$, for reasons given in section~\ref{sec:1ddis}.

The central result is that electronically non-adiabatic RPMD rate theory captures non-adiabatic and quantum effects, even at very low temperatures where RPMD rate theory is expected to be less accurate. Furthermore, the non-adiabatic RPMD rate in Fig~\ref{symrate} is just as good at estimating the non-adiabatic quantum rate, as the adiabatic RPMD rate is at estimating the adiabatic quantum rate. 

\section{Asymmetric curve-crossing}
\subsection{The asymmetric model}

The potential matrix for this model is
\begin{equation}
 {\bf V}(q) = 
\left( \begin{array}{cc}
        Ae^{+Bq} & C        \\
	C        & Ae^{-Bq} - D \\
       \end{array}\right),
\end{equation}
where $A$, $B$ and $C$ retain their previous values and $D = 0.01$ in atomic units. The corresponding adiabatic and diabatic curves are illustrated in Fig~\ref{asymcurve}.

\begin{figure}[h]
\centering
\resizebox{\figwidth}{!} {\includegraphics[angle = 270]{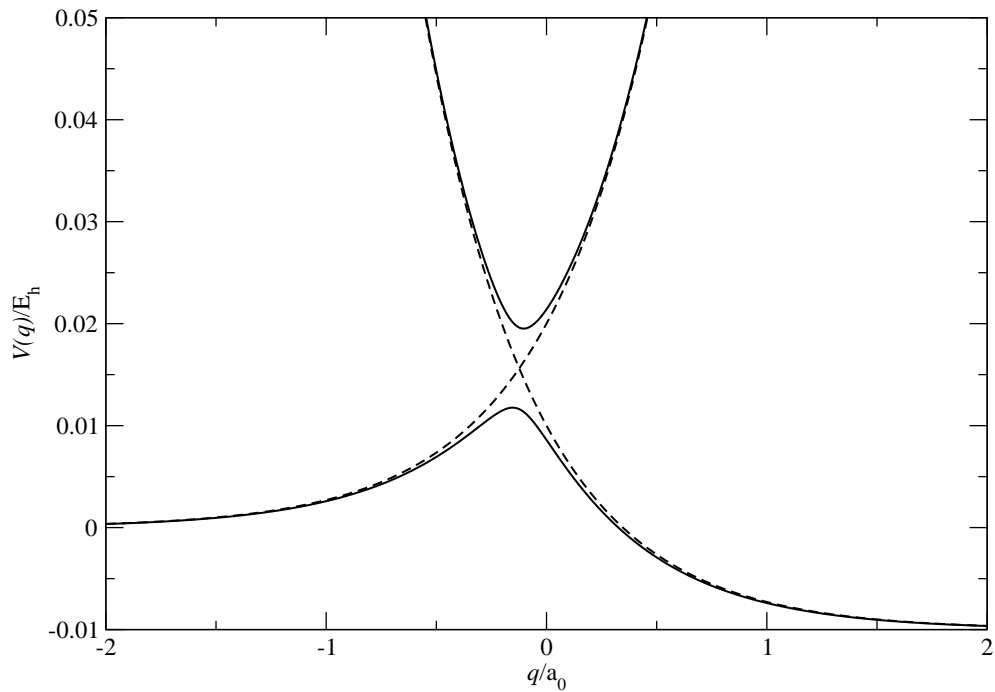}}
\caption{Diabatic (dashed lines) and adiabatic (solid lines) curves for the asymmetric curve crossing model.}
\label{asymcurve}
\end{figure}

Unlike the symmetric model, the optimum dividing surface is no longer apparent by symmetry; although the RPMD rate constant is independent of $q^{\ddag}$ \cite{Mano_cen}, faster convergence is observed with the optimum dividing surface where recrossings are minimised and the transmission coefficient is highest \cite{CH_4}. As discussed in section~\ref{sec:qtst}, the optimum $q^{\ddag}$ can be found by variational minimisation of the QTST rate. 

Exact microcanonical transmission probabilities are calculable as before, and are illustrated in Fig~\ref{neasym}. The transmission probabilities are qualitatively the same, though the adiabatic curve is slightly shifted to lower energies due to a lower barrier height for the reaction. 

\begin{figure}[h]
\centering
\resizebox{\figwidth}{!} {\includegraphics[angle = 270]{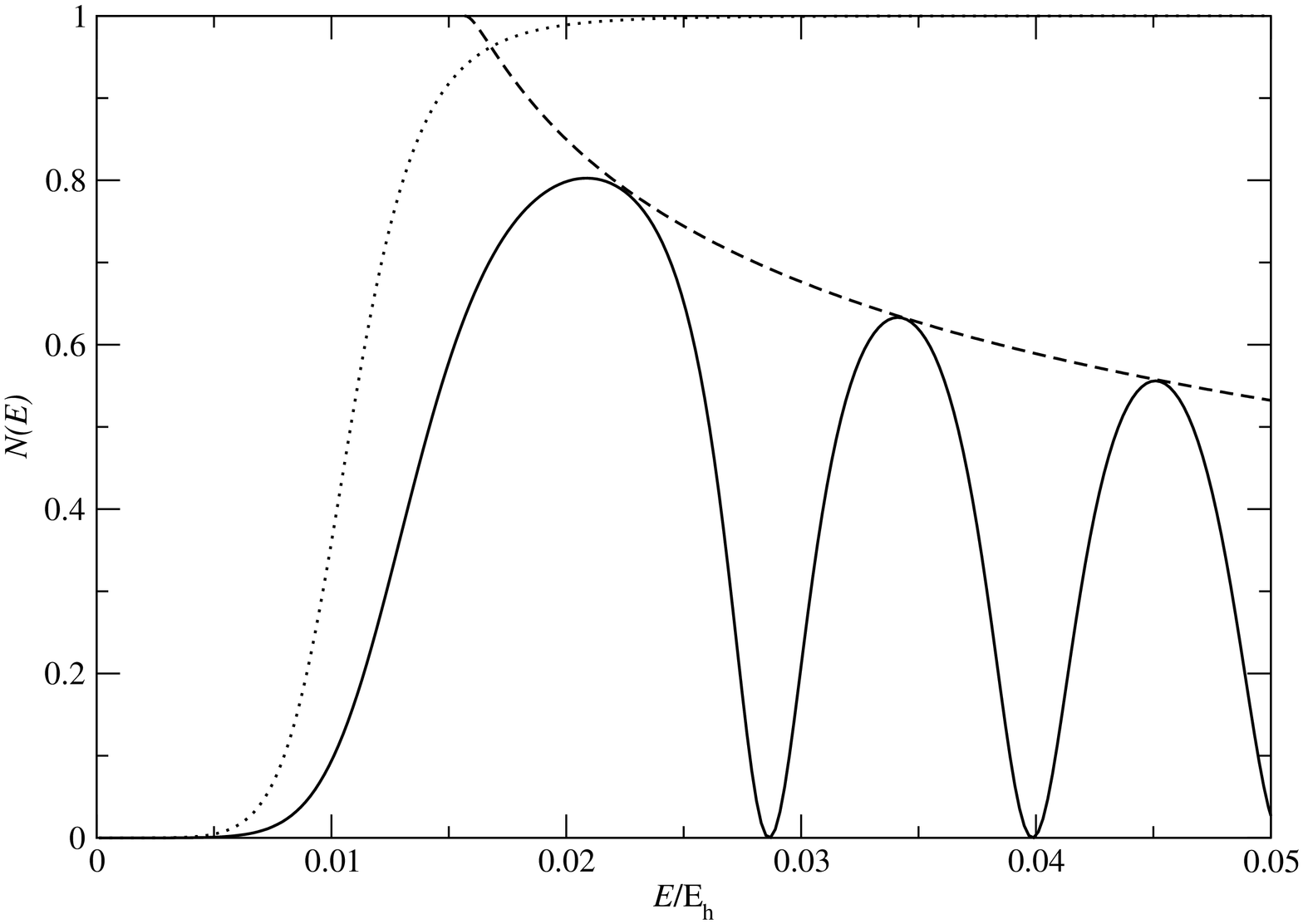}}
\caption{Microcanonical reaction probabilities $N(E)$ for the asymmetric model as a function of microcanonical energy $E$. The dotted line illustrates the reaction probability on the ground adiabatic surface, the dashed line the envelope of the reaction probability arising from a Landau-Zener analysis, and the solid line the reaction probability arising from reaction on the coupled, non-adiabatic surface. The programs for producing these data were provided by D. E. Manolopoulos.}
\label{neasym}
\end{figure}

\subsection{Computational details}
The first step in the asymmetric case is the calculation of the QTST rate for variational optimisation of the dividing surface. From Eq~(\ref{eq:kqtst}),
\begin{equation}
k^{QTST}(T) = \frac{1}{Q_r(T)}\frac{1}{\left(2 \pi \hbar\right)^n}\int d{\bf p} \int d{\bf q} \ e^{-\beta_n H_n({\bf p,q})} \delta(\bar q - q^{\ddag})\frac{\bar p}{m}h\left(\bar p\right).
\end{equation}
The momenta can be integrated out, and using a position-only version of the sampling function in Eq~(\ref{eq:rhoplus}),
\begin{equation}
\rho_{QTST} ({\bf q}) = e^{-\beta_n \frac{1}{2}m\omega_n^2 {\bf q^TAq}} \delta(\bar q - q^{\ddag}),
\end{equation}
the rate becomes
\begin{equation}
k^{QTST}(T) = \frac{1}{\sqrt{2\pi\beta m}} 
\frac{\int d{\bf q}\ \rho_{QTST}({\bf q}) e^{-\beta_n V_n({\bf q})}}{\int d{\bf q} \ \rho_{QTST} ({\bf q})}.
\label{eq:calkqtst}
\end{equation}
Consequently, for calculation of the QTST rate, momenta need not be calculated explicitly. The same position sampling function is used as for the symmetric reaction given in subsection~\ref{sssec:mc}, from which $V_n({\bf q})$ is calculated. $e^{-\beta_n V_n({\bf q})}$ contributions from each sampling of co-ordinates are summed, and their average (multiplied by the $1/\sqrt{2\pi\beta m}$ prefactor) is the QTST rate. This is calculated for a number of different $\bar q = q^{\ddag}$ values, which, using a Golden Section Search \cite{num_rec}, converge upon the optimum value. For each value of $q^{\ddag}$, sampling is iterated to convergence. 

Once the optimum dividing surface is known for each temperature at which the rate is to be calculated, the rate calculation proceeds in the same manner as for the symmetric case, detailed in subsection \ref{subsec:comp}. 

In the `classical' one bead adiabatic case, the optimum dividing surface is independent of temperature and can be calculated analytically as\footnote{This is achieved by calculating the value of $q$ for which the energy of the lower adiabatic electronic eigenstate of ${\bf V}({q})$ is maximized.}
\begin{equation}
q_{cl}^{\ddag} =  \frac{1}{B}{\rm arcsinh}\left(\frac{-D}{2(A-C)}\right).
\end{equation}
The classical rate calculation is therefore analytic for the adiabatic surface. However, since the temperature of the energy gap between the two adiabatic states is so large, it is a reasonable approximation to use the adiabatic dividing surface in the one-bead non-adiabatic rate calculation (and any deficiencies with the choice of $q^{\ddag}$ will be corrected by running the trajectories to sufficiently long time). As in the symmetric case, the adiabatic and non-adiabatic one-bead rates are identical to within graphical accuracy and so only the non-adiabatic rate is plotted.

Convergence parameters were the same as those for the symmetric calculation in subsection~\ref{sssec:conv}.

\subsection{Results}
%
%
Thermal reaction rates for the asymmetric model are presented in Fig~\ref{asymrate}. 
\begin{figure}[h]
\centering
\resizebox{\figwidth}{!} {\includegraphics[angle = 270]{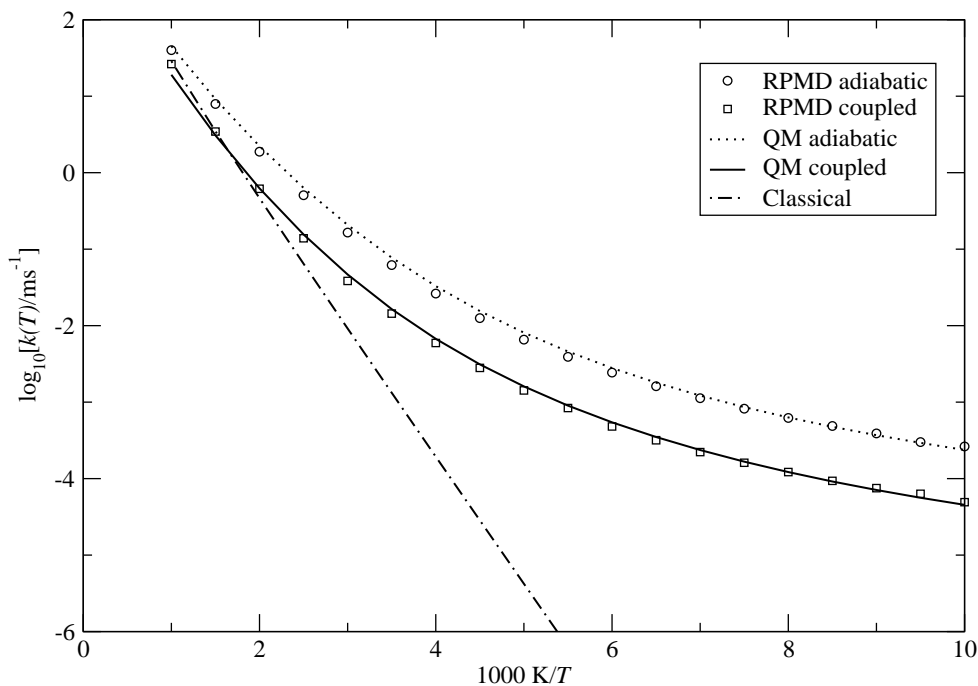}}
\caption{Reaction rates for the asymmetric curve crossing model.}
\label{asymrate}
\end{figure}

The RPMD rate is an even better approximation to the exact quantum rate in the asymmetric model than for the symmetric one. 
The adiabatic rate is greater than the non-adiabatic rate for the same reason as for the symmetric case (Landau-Zener-St\"uckelburger oscillations), and the classical rate again has an Arrhenius temperature dependence. Note that, for a given temperature, the quantum exact and RPMD rates for the asymmetric model are considerably higher than their analogues in the symmetric model due to a lower barrier height, as visible from comparing Figs~\ref{symcurve} and \ref{asymcurve}.


\section{Discussion}
\label{sec:1ddis}
The previous two model problems suggest that the non-adiabatic generalization of RPMD rate theory is a good approximation to exact quantum mechanical rates, even in deep tunnelling where classical rate theory is in error by many orders of magnitude. The error of the non-adiabatic RPMD theory compared to exact quantum results is similar to that of the adiabatic theory, as shown in Figs \ref{symrate} and \ref{asymrate}. The use of the Hamiltonian derived in chapter~\ref{chap:nag} therefore seems to capture the non-adiabatic inhibition of the rate, at least for these one-dimensional curve crossing models.

In deep tunnelling (the low temperature limit) the reaction rate is underestimated for a symmetric potential and overestimated for an asymmetric potential, in accordance with the observations of Richardson and Althorpe for a single potential energy surface \cite{althorpe}. From my results, this appears to apply to reactions on both the adiabatic and electronically coupled potential energy surfaces, even though Richardson et.\ al.\ only considered an adiabatic surface.

In the high temperature limit, both the adiabatic RPMD rate and adiabatic quantum rate tend to the (exact) classical rate, as visible in Figs~\ref{symrate} and \ref{asymrate} and explained in Ref \cite{Mano_pin}. However, the same figures demonstrate that this is not true for the non-adiabatic RPMD rate, which does not become exact in the high temperature limit. As $T \to \infty$, the number of beads required falls such that $n \to 1$ and adiabatic RPMD rate theory reduces to classical rate theory, which (in the high temperature regime) is exact \cite{Mano_pin}. However, there is no high-temperature classical analogue of non-adiabatic RPMD theory, since the non-adiabaticity is an inherently quantum mechanical effect. Consequently, although there is a one-bead limit of non-adiabatic RPMD theory, explored in section \ref{sec:1blimit}, this does not correspond to any exact classical or quantum rate.

The models considered so far are one-dimensional. It is known that adiabatic RPMD rate theory can be extended easily to multidimensional models \cite{Mano_cen}, and in the next chapter we explore the application of the non-adiabatic generalization to a more complex multidimensional model, the spin-boson model \cite{sb}.
\chapter{The Spin-Boson Model}
\label{chap:sb}
In this chapter non-adiabatic RPMD rate theory is extended to a multidimensional application, the spin-boson model. This is a model for quantum mechanical dissipative processes, where a two-level system interacts bilinearly with a harmonic bath \cite{sb}. It has numerous applications, from electron transfer in biomolecules and solution \cite{garg, bio, elec_trans} to quantum computing \cite{qcomp} and wavepacket dynamics \cite{wavepacket}, and a number of approximate \cite{garg, niba} and numerically exact \cite{sbprev} methods exist for its description. 

After introducing the model in section~\ref{sec:sbmodel}, non-adiabatic RPMD rate theory is generalized to multidimensional systems, followed by introducing the Bennett-Chandler factorization \cite{bc} as a practical way to calculate the RPMD rate.  We find that the Quantum Transition State Theory (QTST) rate is undefined for the spin-boson model with Debye spectral density, but by using an unusual factorization of the RPMD rate coefficient a converged RPMD rate is calculable, and preliminary calculations are presented as a function of the coupling strength between the spin system and the harmonic bath.

\section{The model}
\label{sec:sbmodel}
In this section the model is defined by its Hamiltonian, followed by exploring the spectral density and the mass of the reaction co-ordinate. I also outline the exact results used for comparison. 
\subsection{The spin-boson Hamiltonian}
\label{subsec:sbham}
The spin-boson Hamiltonian for a system of $f$ bath modes consists of three components \cite{sb},
\begin{equation}
 H = H_s + H_{b} + H_{sb},
\end{equation}
where $H_s$ is the spin Hamiltonian,
\begin{equation}
 H_s = \epsilon \sigma_z + \Delta \sigma_x,
\end{equation}
$H_{b}$ corresponds to a bath of $f$ Harmonic oscillators,
\begin{equation}
 H_{b} = \sum_{k=1}^f \frac{1}{2}\left(p_k^2 + \omega_k^2 q_k^2 \right),
\label{eq:harmbath}
\end{equation}
and $H_{sb}$ is the system-bath coupling,
\begin{equation}
 H_{sb} = \sigma_z \sum_{k=1}^f c_k q_k.
\end{equation}
Here $\sigma_z$ and $\sigma_x$ are the Pauli spin matrices,
\begin{equation}
 \sigma_z = 
\left(
\begin{array}{cc}
 1 & 0  \\
 0 & -1 
\end{array}
\right),
\qquad
\sigma_x = 
\left(
\begin{array}{cc}
0 & 1 \\
1 & 0
\end{array}
\right).
\end{equation}
Each bath mode has momentum $p_k$, position $q_k$, and a characteristic frequency $\omega_k$ and weight $c_k$. The parameter $\Delta$ controls the coupling between the diabatic states, and the mass $m$ is taken to be unity (in atomic units). We shall consider the symmetric model with zero energy bias, such that $\epsilon = 0$. 

The bath is completely defined by its spectral density,
\begin{equation}
 J(\omega) = \frac{\pi}{2} \sum_{k=1}^f \frac{c_k^2}{\omega_k} \delta(\omega-\omega_k),
\label{eq:specdensum}
\end{equation}
which is a discrete approximation to a continuous density of states (explored in subsection \ref{subsec:dsd}). It will also prove useful to define a reaction co-ordinate $y$, where
\begin{equation}
 y = \sum_{k=1}^f c_k q_k,
\label{eq:ydef}
\end{equation}
simplifying the overall Hamiltonian to
\begin{equation}
 H = \left(
\begin{array}{cc}
 y & \Delta  \\
 \Delta & -y 
\end{array}
\right)
+ \sum_{k=1}^f \frac{1}{2}\left(p_k^2 + \omega_k^2 q_k^2 \right).
\end{equation}
This so-called `Marcus' reaction co-ordinate \cite{marcus}, $y$, clearly corresponds to (half) the energy difference between the two diabatic states. These electronic states are mixed by an amount $\Delta$ from $H_s$ to which harmonic contributions from $H_b$ are added. Adiabatic and diabatic curves for a one-dimensional spin-boson model are illustrated in Fig~\ref{fig:potsurf}.
\newlength{\figwidthf}%
\setlength{\figwidthf}{\columnwidth}%
\begin{figure}[htb]%
\centering%
\resizebox{\figwidthf}{!} {\includegraphics[angle = 270]{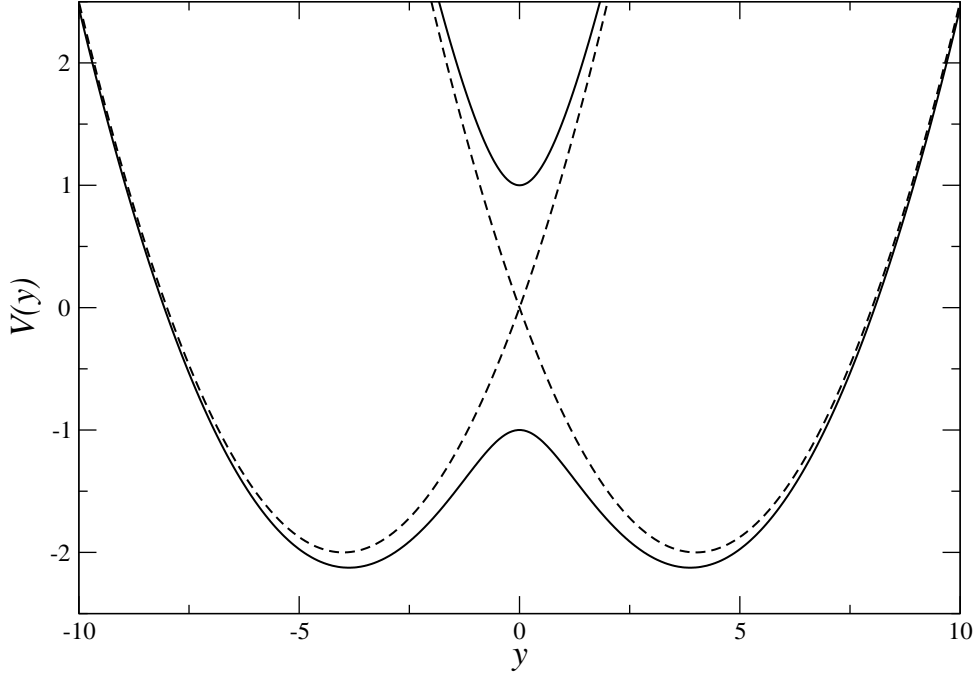}}%
\caption{Illustrative one-dimensional potential for the spin-boson model, where $\Delta=1,\ \omega = \tfrac{1}{2}$ (all parameters in atomic units). Dashed lines represent diabatic surfaces and solid lines adiabatic surfaces.}%
\label{fig:potsurf}%
\end{figure}%
The mixing of the two parabolic diabats leads to a double-well potential, the rate of transfer between the two wells being the reaction rate. The optimum choice of dividing surface for this reaction co-ordinate is evident by symmetry: $y^{\ddag}=0$.
%
\subsection{Debye spectral density}
\label{subsec:dsd}
The continuous spectral density I have chosen to focus on is the so-called Debye type for which exact reaction rates are available \cite{sb},
\begin{equation}
 J(\omega) = \frac{\eta \omega_c \omega}{\omega_c^2 + \omega^2},
\label{eq:debye}
\end{equation}
where $\eta$ is the coupling strength and $\omega_c$ is the characteristic frequency of the bath. This $J(\omega)$ is linear at low frequencies, peaks at $\omega_c$ and has a long, Lorentzian tail at high $\omega$, depicting a solvent with Debye dielectric relaxation \cite{sb}. Its only defined power moment is $J(\omega)/\omega$, which is related to the classical reorganization energy \cite{sbprev},
\begin{equation}
 E_r = \frac{4}{\pi}\int_0^{\infty} d\omega\frac{J(\omega)}{\omega}.
\label{eq:er}
\end{equation}
Substituting $\tan\theta = \omega/\omega_c$ into Eq~(\ref{eq:er}) gives
\begin{equation}
 E_r = \frac{4}{\pi}\int_0^{\infty} \frac{\eta\omega_c}{\omega_c^2 + \omega^2} d\omega = \frac{4}{\pi}\eta\int_0^{\pi/2}d\theta = 2\eta.
\label{eq:er2}
\end{equation}
All sensible discretization schemes should produce the same result in the limit of an infinite number of bath modes, but the aim is to choose one which requires the least number of bath modes, thereby maximizing computational efficiency \cite{markland, miller_semiclas}.

Multiplying the spectral density in Eq~(\ref{eq:debye}) by a generalised moment which I choose to write $F(\omega)/\omega$, and integrating,
\begin{eqnarray}
 \int_0^{\infty} F(\omega)\frac{J(\omega)}{\omega} d\omega & = & \int_0^{\infty} F(\omega)\frac{\eta\omega_c}{\omega_c^2 + \omega^2} d\omega\\
& = & \eta \int_0^{\pi/2} F(\omega_c\tan\theta)d\theta,
\end{eqnarray}
where I have substituted $\tan\theta = \omega/\omega_c$ as in Eq~(\ref{eq:er2}). Using the midpoint rule to approximate the integral into $f$ windows, centered on $\theta_k = (k - \tfrac{1}{2})\pi/2f$,
\begin{equation}
\int_0^{\infty} F(\omega)\frac{J(\omega)}{\omega} d\omega \simeq \frac{\pi\eta}{2f} \sum_{k=1}^f F(\omega_c\tan\theta_k). 
\label{eq:sdsum}
\end{equation}
However, if we multiply Eq~(\ref{eq:specdensum}) by $F(\omega)/\omega$ and integrate,
\begin{eqnarray}
 \int_0^{\infty} F(\omega) \frac{J(\omega)}{\omega} d\omega & = & 
\frac{\pi}{2} \int_0^{\infty} \sum_{k=1}^f F(\omega)\frac{c_k^2}{\omega_k^2} \delta(\omega-\omega_k) d\omega  \\
& = & \frac{\pi}{2} \sum_{k=1}^f \frac{c_k^2}{\omega_k^2}F(\omega_k).
\label{eq:specint}
\end{eqnarray}
Comparing equations~(\ref{eq:sdsum}) and (\ref{eq:specint}) strongly suggests the discretization scheme
\begin{equation}
 \omega_k = \omega_c \tan\left(\frac{(k - \tfrac{1}{2})\pi}{2f}\right),
 \label{eq:debomega}
\end{equation}
and from Eq~(\ref{eq:sdsum}),
\begin{equation}
 c_k = \sqrt{\frac{\eta}{f}} \omega_k,
\end{equation}
where 
\begin{equation}
 k = 1, 2, \ldots, f.
\end{equation}

This discretization scheme exactly calculates the only defined moment of the Debye spectral density for any number of bath modes, and reproduces the Debye spectral density's long frequency tail (the slow decay of $J(\omega)$ as $\omega \to \infty$). This is apparent from the maximum frequency,
\begin{equation}
 \omega_{max} = \omega_c \tan \left( \frac{\pi}{2} - \frac{\pi}{4f} \right) = \omega_c \cot\left(\frac{\pi}{4f}\right) \simeq \frac{4f\omega_c}{\pi},
\label{eq:omegamax}
\end{equation}
which is roughly equal to $10 \omega_c$ for ten bath modes. The absence of convergent positive moments of $J(\omega)$ is a consequence of the slow cutoff, $\lim_{\omega \to \infty} J(\omega) \sim 1/\omega$, which is absent with the commonly used, normalizable Ohmic spectral density, $J(\omega) = \eta\omega e^{-\omega/\omega_c}$, where $\omega_{max} \simeq 3\omega_c$ for 10 bath modes (using the discretization scheme in Ref.~\cite{Mano_pin}). The Ohmic spectral density has been applied to the spin-boson model \cite{sbprev, makri} but is presented here for comparative purposes only.
\subsection{The mass factor}
As stated in subsection~\ref{subsec:sbham}, the mass of a particle in a particular bath mode is chosen to be unity. However, the mass of the reaction co-ordinate itself, $\mu$, is \cite{h_mu_diffusion}
\begin{eqnarray}
 \mu & = & \left(\sum_{k=1}^f \left| \frac{dy}{dq_k} \right|^2 \right)^{-1} \\
& = & \left(\sum_{k=1}^f c_k^2 \right)^{-1} \\
& = & \frac{f}{\eta\omega_c^2 }\left(\sum_{k=1}^f \tan^2\left(\frac{(k - \tfrac{1}{2})\pi}{2f}\right)\right)^{-1}.
\end{eqnarray}
Noting the intriguing identity, whose proof is detailed in appendix~\ref{a:massfac},
\begin{equation}
 \sum_{k=1}^f\tan^2\left(\frac{(k - \tfrac{1}{2})\pi}{2f}\right) = f(2f-1),
\label{eq:massfac}
\end{equation}
the mass factor has the analytic form
\begin{equation}
 \mu = \frac{1}{\eta\omega_c^2 (2f-1)}.
\label{eq:debmass}
\end{equation}
As such, the mass factor decays to zero as the number of bath modes $f$ tends to infinity. This crucial result is a consequence of the nature of the Debye spectral density, and not due to RPMD rate theory, as the mass factor is derived independently of any ring polymer considerations. 
Conversely, the Ohmic spectral density can be discretized \cite{Mano_pin} producing a convergent mass factor $\mu_{f \to \infty} = \pi/4\eta\omega_c^2$. Both mass factors are illustrated in Fig~\ref{fig:massfac}.
\begin{figure}[htb]
\centering
\resizebox{\figwidthf}{!} {\includegraphics[angle = 270]{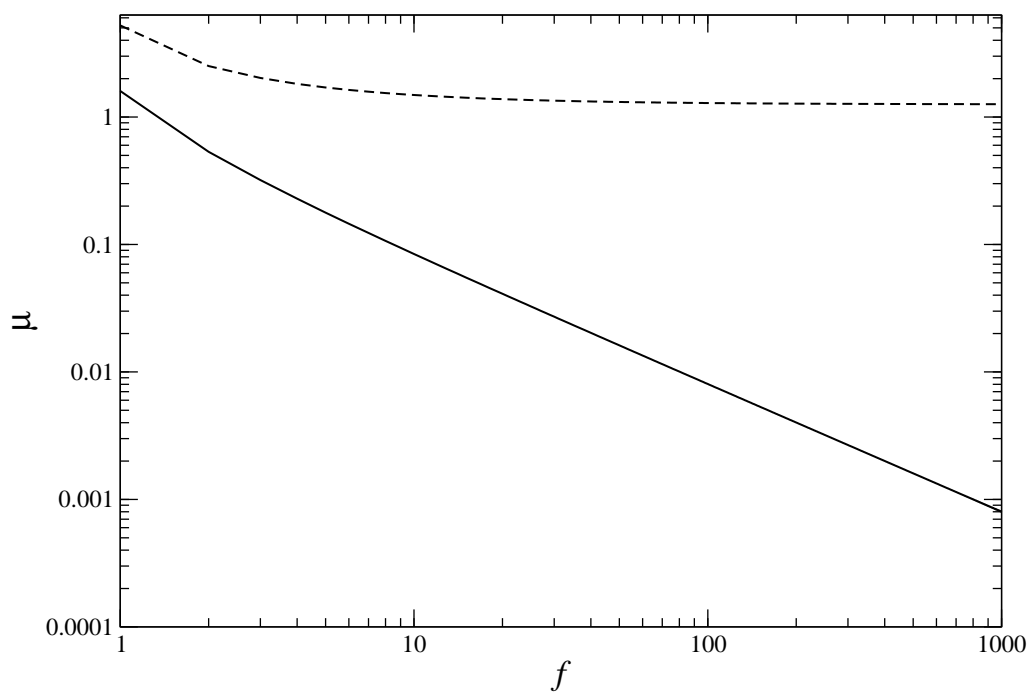}}
\caption{Mass factor $\mu$ for the Debye (solid line) and Ohmic (dashed line) spectral densities as a function of the number of bath modes $f$ with $\eta = 10$ and $\omega_c = \tfrac{1}{4}$ in both cases (all parameters are in atomic units). The Ohmic mass factor is convergent, whereas the Debye mass factor decays as $\mu \sim 1/f$ as $f \to \infty$.}
\label{fig:massfac}
\end{figure}
\subsection{Exact Results}
This version of the spin-boson model with the Debye spectral density is used here because Wang et.\ al.\ have calculated exact quantum rates for this model using a numerically convergent multiconfiguration time-dependent Hartree (MCTDH) method \cite{sb}. The parameters they used, and which are considered here, are $\beta\Delta = 0.5$, $\Delta/\omega_c =4$, and $\eta/\Delta$ ranges from 5 to 15. Their parameters are given relative to $\Delta$, such that they are dimensionless, and here (for convenience) $\Delta=1$ in atomic units. Furthermore, Wang et.\ al.\ compared their results with those of other commonly used approximate methods for the spin-boson model, including the Generalized Smoluchowski Equation \cite{garg}, quantum \cite{weiss} and classical golden rules \cite{marcus}, and the so-called Zusman rate \cite{zusman}. Employing RPMD rate theory with these parameters allows exact results for comparison as well as a benchmark for the accuracy of our calculations compared to other approximate methods. The spin-boson model should therefore be an instructive test for the first application of electronically non-adiabatic RPMD rate theory to a multidimensional model.
\section{Multidimensional generalization of RPMD}%
\label{sec:multidimen}%
Here I extend RPMD rate theory derived in chapters \ref{chap:rpmd} and \ref{chap:nag} to multidimensional systems, in the context of the spin-boson model. This has already been performed for systems with a single electronic potential energy surface \cite{Mano_cen}, and in this section includes the non-adiabatic generalization presented in chapter~\ref{chap:nag}.

Consider a general $f$-dimensional Hamiltonian,
\begin{equation}
 H = \sum_{k=1}^f \tfrac{1}{2}p_k^2 + V(q_1, \ldots, q_f),
\end{equation}
and a dividing surface $s(q_1, \ldots, q_f) = 0$, for which products are found in $s > 0$. 

The ring-polymer flux-side correlation function of Eq~(\ref{cfs_rpmd}) with $n$ beads is now written in terms of the centroids of the ring polymer in each dimension \cite{Mano_cen},
\begin{equation}
c_{fs}(t) = \frac{1}{(2\pi\hbar)^{nf}} \int d{\bf p_0} \int d{\bf q_0} \ e^{-\beta_n H_n({\bf p_0,q_0})} \delta[\bar s({\bf q_0})]\bar v({\bf p_0,q_0}) h[\bar s({\bf q_t})].
\label{eq:cfsmulti}
\end{equation}
Here the delta and heaviside functions are written in terms of $s(\bar q_1, \ldots, \bar q_f) \equiv \bar s({\bf q})$ where 
\begin{equation}
 \bar q_k = \frac{1}{n} \sum_{j=1}^n q_{j,k}.
\end{equation}
Variables $\bf p$ and $\bf q$ are now matrices of momenta and positions, such that $p_{j,k}$ is the momentum of the $j$th bead moving in the $k$th dimension of the ring polymer phase space, and likewise for a co-ordinate $q_{j,k}$. The velocity relative to the dividing surface is \cite{Mano_cen}, 
\begin{equation}
 \bar v({\bf p,q}) = \sum_{k=1}^f \frac{\partial \bar s ({\bf q})}{\partial \bar q_k} \frac{\bar p_k}{m_k},
\label{eq:vbar}
\end{equation}
where the momentum centroid in the $k$th dimension is
\begin{equation}
\bar p_k = \frac{1}{n} \sum_{j=1}^n p_{j,k},
\end{equation}
and $m_k$ is the mass of particle $k$.
The general multidimensional ring polymer Hamiltonian for a single electronic potential energy surface is \cite{Mano_cen}
\begin{equation}
 H_n({\bf p,q}) = \sum_{k=1}^f\sum_{j=1}^n \left[ \frac{p_{j,k}^2}{2m_k} + \tfrac{1}{2}m_k\omega_n^2(q_{j,k}-q_{j+1,k})^2 \right] + \sum_{j=1}^n V(q_{1,j},\ldots,q_{f,j}).
\label{eq:hmulti}
\end{equation}

As detailed in section \ref{sec:effpot}, an electronically non-adiabatic potential cannot be expressed as a sum of the potentials of different beads, but is a nonlinear function of all positions, such that $\sum_{j=1}^n V(q_{1,j},\ldots,q_{f,j})$ becomes $V_n({\bf q})$, where
\begin{equation}
 V_n({\bf q}) = - \frac{1}{\beta_n} \log \left[\sum_{j_1} \ldots \sum_{j_n} \langle j_1 | e^{-\beta_n \hat V(q_{1,1},\ldots,q_{f,1})}| j_2 \rangle \ldots \langle j_n |e^{-\beta_n \hat V(q_{1,n},\ldots,q_{f,n})}| j_1 \rangle \right],
\label{eq:vlong}
\end{equation}
which is a multidimensional generalisation of the non-adiabatic potential in Eq~(\ref{eq:napotential}). Note that $j_i$ values refer to electronic states, $j$ values to bead numbers. The time-evolved side operator $h[\bar s({\bf q_t})]$ is calculated by evolving the ring polymer through an extended classical phase space under the influence of the Hamiltonian in Eq~(\ref{eq:hmulti}) with the potential in Eq~(\ref{eq:vlong}).

The formulae presented in Eqs~(\ref{eq:cfsmulti}) to (\ref{eq:vlong}) are applicable to any multidimensional system with a non-adiabatic potential. However, for the spin-boson model, both the flux-side correlation function in Eq~(\ref{eq:cfsmulti}) and the potential in Eq~(\ref{eq:vlong}) can be simplified significantly. 

From the definition of a reaction co-ordinate and a dividing surface in section~\ref{subsec:sbham},
\begin{equation}
 s(q_1, \ldots, q_f) \equiv \sum_{k=1}^f c_k q_k \equiv y, 
\end{equation}
and
\begin{equation}
 \frac{\partial \bar s ({\bf q})}{\partial \bar q_k} = c_k.
\end{equation}
Consequently,
\begin{eqnarray}
 \bar s({\bf q}) & = & \sum_{k=1}^f c_k \bar q_k \nonumber \\
& = & \sum_{k=1}^f c_k \left (\frac{1}{n}\sum_{j=1}^n q_{j,k}\right) \nonumber \\
& = & \frac{1}{n}\sum_{j=1}^n y_j  \nonumber \\
& = & \bar y,
\end{eqnarray}
where the reaction co-ordinate centroid is defined as
\begin{equation}
 \bar y = \frac{1}{n} \sum_{j=1}^n y_j. 
\end{equation}
Choosing to define a reaction co-ordinate momentum centroid as
\begin{equation}
\bar p_y = \mu \sum_{k=1}^f c_k \bar p_k,
\end{equation}
where $\mu$ is the mass of the reaction co-ordinate, and noting that $m_k=1$, simplifies Eq~(\ref{eq:vbar}),
\begin{equation}
 \bar v({\bf p,q}) = \sum_{k=1}^f c_k \bar p_k = \frac{\bar p_y}{\mu}.
\end{equation}
Applying these notational simplifications to Eq~(\ref{eq:cfsmulti}) leads to a more succinct flux-side correlation function for the spin-boson model,
\begin{equation}
 c_{fs}(t) = \frac{1}{(2\pi\hbar)^{nf}} \int d{\bf p_0} \int d{\bf q_0} \ e^{-\beta_n H_n({\bf p_0,q_0})} \delta(\bar y_0)\frac{\bar p_y}{\mu} h(\bar y_t).
\end{equation}
Furthermore, for the spin-boson model, the potential for a single particle is separable into an electronically non-adiabatic part and a part which is independent of electronic state occupancy,
\begin{equation}
 \hat V({\bf q}) = \hat V^{elec}({\bf q}) + V^{0}({\bf q}),
\end{equation}
 where
\begin{equation}
 \hat V^{elec}({\bf q}) = H_s + H_{sb} = y \sigma_z + \Delta \sigma_x = 
\left(
\begin{array}{cc}
 y & \Delta  \\
 \Delta & -y 
\end{array}
\right),
\end{equation}
and
\begin{equation}
 V^{0}({\bf q}) = \sum_{k=1}^f \frac{1}{2}\omega_k^2 q_k^2.
\end{equation}
As seen in subsection~\ref{subsec:mixp}, only the electronically non-adiabatic part of the potential $\hat V^{elec}({\bf q}) $ need be considered using the complicated equation~(\ref{eq:napotential}), whereas $V^{0}({\bf q})$ can be expressed as a simple sum of contributions from each bead. Furthermore, $\hat V^{elec}({\bf q})$ is only a function of bead positions along the reaction co-ordinate $\bf y$ allowing us to write for a ring polymer,
\begin{equation}
 V_n({\bf q}) = V_n^{elec}({\bf y}) + V_n^{0}({\bf q}),
\end{equation}
with
\begin{equation}
 V_n^{elec}({\bf y}) = - \frac{1}{\beta_n} \log \left[\sum_{j_1} \ldots \sum_{j_n} \langle j_1 | e^{-\beta_n \hat V(y_1)}| j_2 \rangle \ldots \langle j_n |e^{-\beta_n \hat V(y_n)}| j_1 \rangle \right],
\label{eq:vy}
\end{equation}
where $\hat V(y_j) = y_j \sigma_z + \Delta \sigma_x$, and 
\begin{equation}
 V_n^{0}({\bf q}) = \sum_{j=1}^n \sum_{k=1}^f\tfrac{1}{2} \omega_k^2 q_{j,k}^2.
\end{equation}
The total ring-polymer Hamiltonian for the spin-boson model can therefore be rewritten as 
\begin{equation}
 H_n({\bf p,q}) = \sum_{j=1}^n \left[\sum_{k=1}^m\tfrac{1}{2} p_{j,k}^2 + \tfrac{1}{2}\omega_n^2(q_{j,k}-q_{j+1,k})^2  + \tfrac{1}{2} \omega_k^2 q_{j,k}^2\right] + V_n^{elec}({\bf y}).
\label{eq:complexh}
\end{equation}
Eq~(\ref{eq:complexh}) can be simplified by using a normal mode transform similar to that detailed in subsection \ref{sssec:nmodes}, 
leading to a Hamiltonian composed of a harmonic ring polymer and an electronic term,
\begin{eqnarray}
 H_n({\bf p,q}) & = & H_n^{hrp}({\bf p,q}) + V_n^{elec}({\bf y}), \label{eq:sbhsplit}\\
\noalign{\hbox{where the harmonic ring polymer Hamiltonian in the normal mode representation is}} \nonumber \\
H_n^{hrp}({\bf \tilde p, \tilde q})& = & \sum_{\tilde \jmath=0}^{n-1} \sum_{k=1}^f\left(\tfrac{1}{2} \tilde p_{\tilde\jmath,k}^2 + \tfrac{1}{2}\tilde\omega_{\tilde\jmath,k}^2\tilde q_{\tilde\jmath,k}^2 \right),
\end{eqnarray}
and $V_n^{elec}({\bf y})$ is given in Eq~(\ref{eq:vy}).
The frequencies in the normal mode representation are, from Eqs~(\ref{eq:omegak}) and (\ref{eq:debomega}),
\begin{equation}
 \tilde \omega_{\tilde\jmath,k} = \sqrt{4\omega_n^2 \sin^2\left(\frac{\tilde\jmath\pi}{n}\right) + \omega_c^2 \tan^2\left(\frac{(k + \tfrac{1}{2})\pi}{2f}\right)},
\label{eq:nmfreq}
\end{equation}
where $\tilde \omega_{\tilde\jmath,k}$ is the frequency of the $\tilde\jmath$th normal mode of a free ring polymer moving in the $k$th mode of the spin-boson bath. 

This multidimensional generalization, though notationally more complex than the one dimensional case, retains many of the desirable features discussed in section~\ref{sec:rpmd}. In the limit of a single bath mode, $f=1$ and Eqs~(\ref{eq:cfsmulti}) and (\ref{eq:hmulti}) reduce to their one-dimensional counterparts, Eqs~(\ref{cfs_rpmd}) and (\ref{eq:hamiltonian}). The flux-side correlation function, from which the rate is calculated, is a classical flux-side correlation function evaluated in $n \times f$ dimensional phase space, and (like the classical and one-dimensional RPMD correlation coefficients) its long-time limit is independent of dividing surface location \cite{Mano_cen}.
\section{The Bennett-Chandler method}
\label{sec:bc}
After generalizing RPMD rate theory to a multidimensional non-adiabatic model, we now explore how a factorization of the RPMD rate constant facilitates its calculation in a bounded potential.
Introducing the notation $\langle \cdots \rangle$ to denote an (unnormalized) thermal average over phase space, 
\begin{equation}
 \langle \cdots \rangle = \frac{1}{(2\pi\hbar)^{nf}} \int d{\bf p_0} \int d{\bf q_0} \ e^{-\beta_n H_n({\bf p_0,q_0})} (\cdots),
\end{equation}
from Eqs~(\ref{eq:kRPMD}) and (\ref{eq:cfsmulti}), the RPMD rate for a general multidimensional reaction can be written as
\begin{equation}
k^{RPMD}(T) = \lim_{t\to\infty} \frac{\langle \delta[\bar s({\bf q_0})]\bar v({\bf p_0,q_0}) h[\bar s({\bf q_t})] \rangle}{\langle h[-\bar s({\bf q_0})] \rangle},
\end{equation}
and \cite{proton_transfer}
\begin{equation}
 Q_{r}(T) = \langle h[-\bar s({\bf q_0})] \rangle.
\end{equation}
The presence of the harmonic bath in $H_b$ (Eq~(\ref{eq:harmbath})) means that as the reaction co-ordinate, $y \to \pm \infty$, the potential energy of the system $V_n({\bf q}) \to \infty$. Consequently, the reactant partition function cannot be that of a free particle, as it is for the barrier transmission problem in chapter \ref{chap:1D}, and its absolute value is, in general, difficult to determine. To circumvent this problem, the RPMD rate constant is multiplied and divided by the thermal positive-momentum flux through the dividing surface, $\langle \delta[\bar s({\bf q_0})]\bar v({\bf p_0,q_0}) h[\bar p_0] \rangle$, leading to \cite{proton_transfer, bc}
\begin{equation}
 k^{RPMD}(T) = \lim_{t\to\infty}\frac{\langle \delta[\bar s({\bf q_0})]\bar v({\bf p_0,q_0}) h[\bar p_0] \rangle}{\langle h[-\bar s({\bf q}_0)] \rangle}\times \frac{\langle \delta[\bar s({\bf q_0})]\bar v({\bf p_0,q_0}) h[\bar s({\bf q_t})] \rangle}{\langle \delta[\bar s({\bf q_0})]\bar v({\bf p_0,q_0}) h[\bar p_0] \rangle}.
\label{eq:bcsplit}
\end{equation}
The first fraction of Eq~(\ref{eq:bcsplit}) is, by comparison with Eq~(\ref{eq:kqtst}), the Quantum Transition State Theory (QTST) rate,
\begin{eqnarray}
 k^{QTST}(T) & = & \frac{\langle \delta[\bar s({\bf q_0})]\bar v({\bf p_0,q_0}) h[\bar p_0] \rangle}{\langle h[-\bar s({\bf q_0})] \rangle} \label{eq:comp}\\
& = & \frac{1}{\sqrt{2\pi\beta\mu}} \frac{\langle  \delta[\bar s({\bf q_0})] \rangle}{\langle h[-\bar s({\bf q_0})] \rangle} \label{eq:simp}.
\end{eqnarray}
The conversion of Eq~(\ref{eq:comp}) to Eq~(\ref{eq:simp}) is achieved by integrating out the momenta. The second fraction in Eq~(\ref{eq:bcsplit}) corresponds to the transmission coefficient,
\begin{equation}
 \kappa(t) = \frac{\langle \delta[\bar s({\bf q_0})]\bar v({\bf p_0,q_0}) h[\bar s({\bf q}_t)] \rangle}{\langle \delta[\bar s({\bf q_0})]\bar v({\bf p_0,q_0}) h[\bar p_0] \rangle}.
\label{eq:transcoef}
\end{equation}
This dimensionless factor accounts for recrossing of the dividing surface \cite{CH_4}; for any system and dividing surface $\kappa(t \to 0_+) = 1$ and $\kappa(t \to \infty) \leq 1$.

Equations (\ref{eq:bcsplit}) to (\ref{eq:transcoef}) are general to any model, but for the spin-boson model further simplifications are possible.
The QTST rate is calculated by noting that the probability of finding the centroid at position $y'$ in the reactant region is \cite{proton_transfer}
\begin{equation}
 P(y') = \frac{\langle  \delta(y'-\bar y) \rangle}{\langle h(y^{\ddag}-\bar y) \rangle},
\label{eq:pden}
\end{equation}
such that 
\begin{equation}
 k^{QTST}(T) = \frac{1}{\sqrt{2\pi\beta\mu}} P(y^{\ddag}).
\label{eq:QTST}
\end{equation}
For this model, by symmetry,
\begin{equation}
 \frac{\langle h(y^{\ddag}-\bar y) \rangle}{\langle 1 \rangle} = \frac{1}{2},
\end{equation}
so Eq~(\ref{eq:QTST}) can be simplified further to
\begin{equation}
  k^{QTST}(T) = \sqrt{\frac{2}{\pi\beta\mu}}\frac{\langle  \delta(y^{\ddag}-\bar y) \rangle}{\langle 1 \rangle}.
\end{equation}
One therefore performs an unconstrained, thermalized simulation to find the probability density of centroids at the dividing surface, which is scaled by $\sqrt{2/\pi\beta\mu}$ to produce the QTST rate. The transmission coefficient is calculated by sampling configurations of the ring polymer with the centroid at the dividing surface, the momenta from the Boltzmann distribution, and evolving the unconstrained ring polymer trajectory to long time, at each time step noting on which side of the dividing surface it lies, and accumulating contributions to the numerator and denominator of Eq~(\ref{eq:transcoef}).
\section{Computational details}
\subsection{QTST calculation}
The QTST rate in Eq~(\ref{eq:simp}) is the canonical probability density of ring polymer position centroids at the dividing surface for a ring polymer moving in the reactant region under the influence of the Hamiltonian $H_n({\bf p,q})$, multiplied by the mean magnitude of the velocity. An (arbitrary) initial configuration of $q_{j,k} = 0\ \forall j,k$ was chosen, and the momenta were sampled from the Boltzmann distribution $e^{-\beta_n p_{j,k}^2/2m}$ contained in $e^{-\beta_n H_n({\bf p,q})}$. 

This configuration was time evolved under the influence of a Langevin thermostat (discussed in subsection \ref{sssec:langevin}) for $10^3$ time units, without contributing to the rate calculation, in order to ensure the ring polymer was fully thermalized. The ring polymer was then evolved for $10^4$ time units using a time step of $0.01$. After each time step, the position of the reaction co-ordinate centroid $\bar y$ was calculated, and contributed to a probability density distribution containing $600$ windows between $-30$ and $+30$ position units. The entire calculation was performed in atomic units, such that 1 time unit $= \hbar/E_h$, where $E_h$ is a Hartree, and position is measured in multiples of the Bohr radius $a_0$. 

For each set of parameters, $2000$ trajectories were run, though in practice convergence was obtained much sooner. Although, for the QTST calculation, only the probability density of finding the ring polymer centroid at the dividing surface is required, plotting probability densities across the whole reaction co-ordinate $\bar y$ allows for an easy test of convergence, since the probability density distribution should be symmetric for a symmetric potential. Furthermore, the free energy surface itself can be calculated as \cite{proton_transfer}
\begin{equation}
 W(y') = -\frac{1}{\beta} \ln P(y'),
\label{eq:freeen}
\end{equation}
where $W(y')$ is the free energy of a ring polymer centred at $y'$, and $P(y')$ is the probability density defined in Eq~(\ref{eq:pden}). The value of the normalized histogram at $y'=0$, multiplied by  $\sqrt{2/\pi\beta\mu}$, is the QTST rate.

\subsubsection{Time evolution}
\label{sssec:te}
A numerical integration scheme similar to that detailed in subsection \ref{sssec:1dte} is used, evolving the momenta under the influence of $V_n^{elec}({\bf y})$ for $\delta t/2$, then momenta and positions with $H_n^{hrp}({\bf p, q})$ for $\delta t$, and finally with $V_n^{elec}({\bf y})$ for $\delta t/2$.
Each ring polymer normal mode $\tilde \jmath$ in each bath mode $k$ has a different evolution frequency $\tilde \omega_{\tilde\jmath,k}$. 

Incorporating the harmonic spin-boson bath modes in $H_n^{hrp}({\bf p,q})$ leads to more accurate trajectories and better energy conservation compared to an algorithm with the same time step which only evolves the normal modes analytically, and places the harmonic bath in the numerically integrated potential. 

In order to calculate the force on each bead, the reaction co-ordinate bead positions $\bf y$ are calculated from the positions $\bf q$, followed by the force on each reaction co-ordinate bead $\partial V_n^{elec}({\bf y})/\partial y_j$. This is converted to the force on each bead $j$ in each mode $k$ by noting that
\begin{equation}
 \frac{\partial V_n({\bf y})}{\partial q_{j,k}} = 
 \frac{\partial V_n({\bf y})}{\partial y_j} \times 
 \frac{\partial y_j}{\partial q_{j,k}} = 
 \frac{\partial V_n({\bf y})}{\partial y_j} \times c_k,
\end{equation}
where we have noted that $\partial y_j/\partial q_{j,k}$ is not a function of $j$ and evaluated it from Eq~(\ref{eq:ydef}). 
\subsubsection{Langevin thermostat}
\label{sssec:langevin}
A variety of thermostats exist for simulating the canonical ensemble \cite{bc}, which should all produce the same result but with different levels of efficiency. As there is such a wide spectrum of frequencies in the spin-boson model with Debye spectral density\footnote{For $f$ bath modes the ratio between the highest and lowest bath frequency modes is approximately $f^2$.}, a Langevin Thermostat tuned to normal modes of the system is used, similar to the path integral Langevin Equation thermostat advocated by Ceriotti et.\ al.\ \cite{michele}. However, instead of using normal modes of the free ring polymer (which leads to an arbitrary friction coefficient of the centroid mode), the harmonic modes of the free ring polymer and bath $\tilde\omega_{\tilde\jmath,k}$ are used, in order to calculate the friction coefficients. Before and after the existing algorithm in subsection \ref{sssec:1dte}, the momenta $\bf p$ are transformed to their harmonic modes $\bf \tilde p$ and then individually thermostatted \cite{michele}, 
\begin{equation}
 \tilde p_{\tilde\jmath,k} \leftarrow c^{(1)}_{\tilde\jmath,k} \tilde p_{\tilde\jmath,k} + \sqrt{\frac{1}{\beta_n}}c^{(2)}_{\tilde\jmath,k} \xi_{\tilde\jmath,k},
\end{equation}
where
\begin{eqnarray}
 c^{(1)}_{\tilde\jmath,k} & = & e^{-\tilde\omega_{\tilde\jmath,k} \delta t}, \\
 c^{(2)}_{\tilde\jmath,k} & = & \sqrt{1-[c^{(1)}_{\tilde\jmath,k}]^2}.
\end{eqnarray}
Here $\delta t$ is the time step, and $\xi_{\tilde\jmath,k}$ a Gaussian deviate with zero mean and unit variance which is different for each normal mode, each bath mode, and each iteration of the algorithm. The thermostatted momenta $\bf \tilde p$ are then transformed back to the bead representation $\bf p$ and the algorithm continues.

\subsubsection{Convergence}
To my surprise, only one bead and ten bath modes were required for convergence of the probability density along the reaction co-ordinate $\bar y$ in the parameter regime $\beta = 0.5, \ \Delta/\omega_c = 4$; this was checked by running simulations with 20 and 40 bath modes, 4 and 16 beads, and noting that the resulting probability densities were identical to within graphical accuracy. The requirement of only one bead suggests that the parameters used, particularly the inverse temperature $\beta$, are in the high-temperature, `classical' limit where only one bead is required. 

Time step convergence of $\delta t = 0.01$ was checked by a simulation with $\delta t=0.001$ producing the same results. An indication that this is a `safe' time step is that, from Eq~(\ref{eq:omegamax}), $\omega_{max} = 10/\pi$, so a time step of 0.01 is roughly 30 times smaller than the period of the fastest bath mode oscillation. Convergence with respect to the number of trajectories is evident from the symmetrical probability densities. Note that the converged probabilities do not equate to a converged $k^{QTST}(T)$, as it omits the $\sqrt{2/\pi\beta\mu}$ factor. This is discussed further in the results, section~\ref{sec:sbresults}.
\subsection{Transmission coefficient calculation}
From Eq~(\ref{eq:transcoef}),
\begin{equation}
 \kappa(t) = \frac{\langle \delta[\bar s({\bf q_0})]\bar v({\bf p_0,q_0}) h[\bar s({\bf q}_t)] \rangle}{\langle \delta[\bar s({\bf q_0})]\bar v({\bf p_0,q_0}) h[\bar p_0] \rangle},
\end{equation}
which, for the spin-boson problem, reduces to
\begin{equation}
 \kappa(t) = \frac{\langle \delta(\bar y_0)\bar p_{y,0} h(\bar y_t) \rangle}{\langle \delta(\bar y_0) \bar p_{y,0} h(\bar p_{y,0}) \rangle},
\label{eq:kappasb}
\end{equation}
where $\bar p_{y,0}$ is the momentum centroid along the reaction co-ordinate at time \mbox{$t=0$}. In order to sample initial configurations where $\bar y_0 = y^{\ddag} = 0$, a thermalized, constrained `mother' simulation is run, from which configurations are randomly sampled approximately once per time unit. From this configuration, momenta are resampled from the Boltzmann distribution and $\bar p_{y,0}$ is calculated. If $\bar p_{y,0} > 0$, it is added to the denominator of the transmission coefficient in Eq~(\ref{eq:kappasb}). An unconstrained, unthermostatted `daughter' trajectory is then run, and after each time step $\bar y$ is calculated. If $\bar y > 0$, $\bar p_{y,0}$ is added to the numerator of the transmission coefficient $\kappa(t)$. Constrained molecular dynamics is achieved using the RATTLE algorithm \cite{rattle}, with constraints $\bar y = 0$ and $\bar p_{y}=0$; time evolution and thermostatting are the same as those used in the QTST calculation, explained in subsections~\ref{sssec:te} and \ref{sssec:langevin}.

As will be elaborated in the results section, converging these calculations presents numerous challenges, and is a matter for future research. 
\section{Preliminary Results}
\label{sec:sbresults}
Here I present the calculations for the QTST rate, and explain why the QTST rate is undefined for the spin-boson model with the Debye spectral density. The transmission coefficient is then examined, from which an alternative factorization of the RPMD rate is proposed.
\subsection{QTST rate}
\label{subsec:probden}
The QTST rate can be expressed as a product of half the thermal magnitude of the velocity and the probability density of centroids at the dividing surface, as detailed in section~\ref{sec:bc}. It was found that the probability density converged with respect to the number of bath modes $f$, and accordingly converged histograms of $P(y)$ for the $\Delta/\omega_c = 4, \ \beta\Delta = 0.5$ regime are illustrated in Fig~\ref{fig:naw4hist}.

As the value of the coupling parameter $\eta/\Delta$ rises from 5 to 15, the stronger interaction between the harmonic bath modes and spin subsystem increases the size of the reaction barrier, thereby decreasing the probability of finding the centroid at the barrier and moving the probability maxima (and potential energy wells) further apart. Nevertheless, even for a coupling parameter as low as $\eta/\Delta = 5$, a double-well potential is still observed. 
\begin{figure}[htb]
\centering
\resizebox{\figwidthf}{!} {\includegraphics[angle = 270]{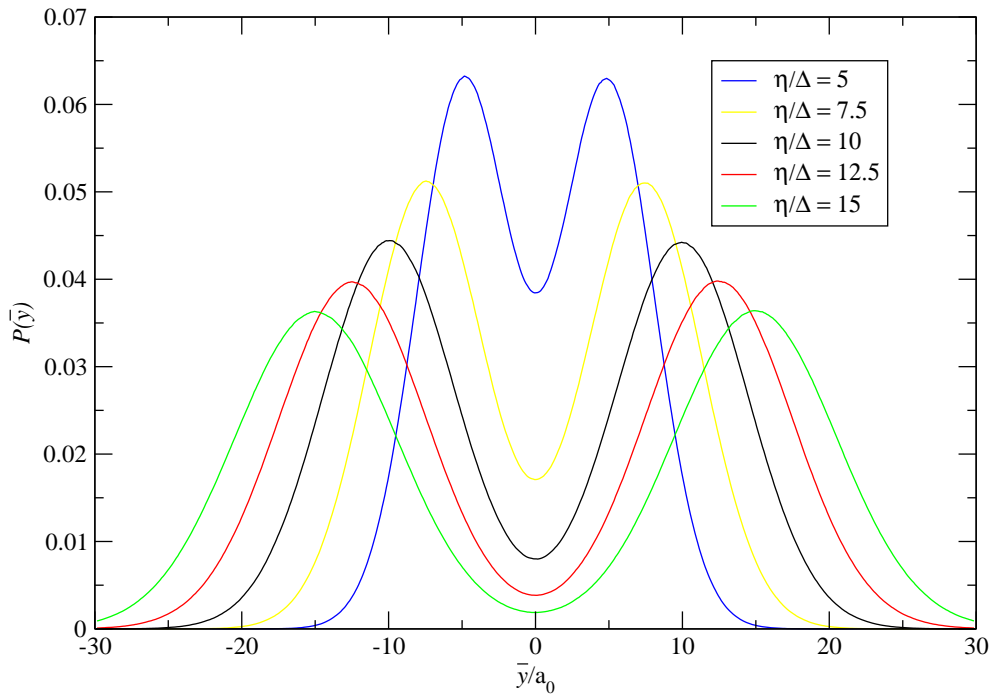}}
\caption{Probability densities for $\beta\Delta = 0.5$, $\Delta/\omega_c = 4$ and varying values of the coupling strength $\eta/\Delta$.}
\label{fig:naw4hist}
\end{figure}

However, defining a QTST rate from these histograms is impossible, as I have so far omitted the prefactor $\sqrt{2/\pi\beta\mu}$. The QTST rate should, {\it prima facie}, be convergent in the limit as the number of modes $f \to \infty$. However, the histograms above are invariant to the number of modes used, and (as illustrated in Fig~\ref{fig:massfac}) $\mu \to 0$ as $f \to \infty$. Consequently, $k^{QTST}(T) \to \infty$ as $f \to \infty$ and the QTST rate is undefined. Physically, the vanishing mass of the reaction co-ordinate suggests a diffusive process (Brownian motion) across the dividing surface \cite{kramers}.

\subsection{Transmission coefficient}
\label{subsec:kappa}
Transmission coefficients for the $\eta/\Delta = 10$ regime are presented in Fig~\ref{fig:transcoef}. In this graph, one bead is used (the `classical' or `high-temperature' limit), though preliminary calculations with 10 and 100 modes and a greater number of beads suggest this is close to convergence.
\begin{figure}[htb]
\centering
\resizebox{\figwidthf}{!} {\includegraphics[angle = 270]{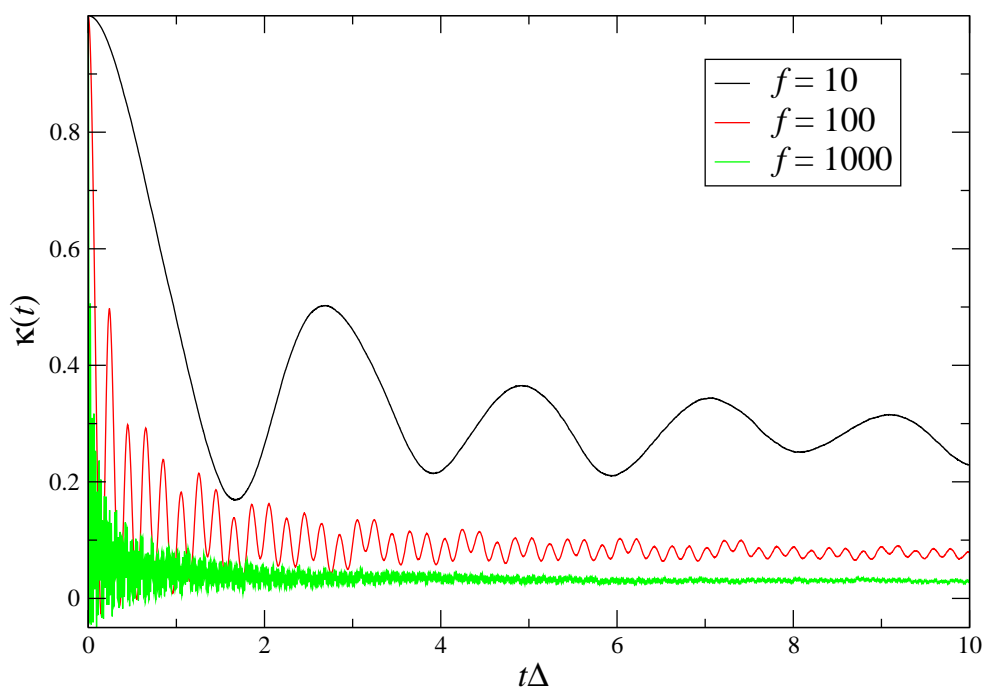}}
\caption{Transmission coefficients for $\beta\Delta = 0.5$, $\Delta/\omega_c = 4$ and $\eta/\Delta = 10$ for varying numbers of bath modes $f$.}
\label{fig:transcoef}
\end{figure}

As the number of bath modes rises, the transmission coefficient falls, and the oscillations in the transmission coefficient decrease in amplitude and increase in frequency. For a low number of bath modes, no plateau appears, which is required to define a rate \cite{varopt}. Furthermore, it appears that the transmission coefficient does not converge with respect to the number of bath modes, such that defining a converged transmission coefficient is impossible.

The observed oscillations are probably due to the high frequency bath modes, which, in this extended classical phase space, oscillate classically in the bottom of their wells. In a true `quantum' simulation, these modes would almost always be in their ground state and probably not cause the observed fluctuations in the transmission coefficient\footnote{For $f=100$ and the Debye spectral density, $\hbar \omega_{max} > 15k_BT$.}. If a spectral density with a faster cutoff (such as the Ohmic spectral density) were used, there would not be as many high frequency modes and the large oscillations would be less likely to occur.

\subsection{Alternative factorization}%
\label{subsec:altfac}
From subsections \ref{subsec:probden} and \ref{subsec:kappa}, the QTST rate tends to infinity as the number of bath modes rises, and the transmission coefficient tends to zero. This suggests that, combining them, one may be able to define an overall rate, $k^{RPMD}(T)$, if --- and only if --- the divergence to infinity and vanishing to zero cancel each other. From Eq~(\ref{eq:bcsplit}), I have defined
\begin{equation}
 k^{RPMD}(T) = \underbrace{\frac{\langle \delta[\bar s({\bf q_0})]\bar v({\bf p_0,q_0}) h[\bar p_0] \rangle}{\langle h[-\bar s({\bf q_0})] \rangle}}_{k^{QTST}(T)}
\times \lim_{t\to\infty}
\underbrace{\frac{\langle \delta[\bar s({\bf q_0})]\bar v({\bf p_0,q_0}) h[\bar s({\bf q_t})] \rangle}{\langle \delta[\bar s({\bf q_0})]\bar v({\bf p_0,q_0}) h[\bar p_0] \rangle}}_{\kappa(t)}.
\end{equation}
However, from Eq~(\ref{eq:simp}) one could alternatively define the rate as 
\begin{equation}
 k^{RPMD}(T) =  \underbrace{\frac{\langle  \delta[\bar s({\bf q_0})] \rangle}{\langle h[-\bar s({\bf q_0})] \rangle}}_{P(y^{\ddag})} 
\times \lim_{t\to\infty} \underbrace{\frac{1}{\sqrt{2\pi\beta\mu}} \frac{\langle \delta[\bar s({\bf q_0})]\bar v({\bf p_0,q_0}) h[\bar s({\bf q}_t)] \rangle}{\langle \delta[\bar s({\bf q_0})]\bar v({\bf p_0,q_0}) h[\bar p_0] \rangle}}_{\tfrac{1}{2}\langle | \dot y | \rangle \kappa(t)}.
\label{eq:altfac}
\end{equation}
From Fig~\ref{fig:naw4hist}, $P(y^{\ddag})$ is converged with respect to the number of modes, suggesting calculation of $\tfrac{1}{2}\langle | \dot y | \rangle \kappa(t)$. This is illustrated in Fig~\ref{fig:modemass}. 
\begin{figure}[htb]
\centering
\resizebox{\figwidthf}{!} {\includegraphics[angle = 270]{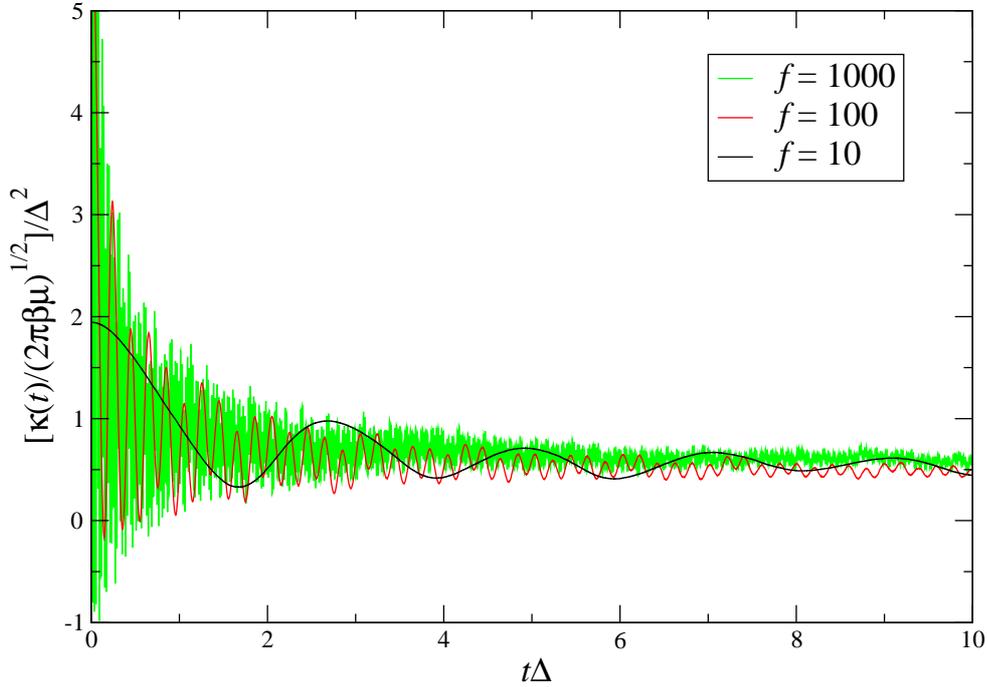}}
\caption{Transmission coefficients from Fig~\ref{fig:transcoef} multiplied by the mean magnitude of the velocity, for varying values of $f$. Note that, in the long time limit, the transmission coefficients for different $f$ appear to converge upon the same value.}
\label{fig:modemass}
\end{figure}

From these preliminary calculations, it appears that using this factorization, $\tfrac{1}{2}\langle | \dot y | \rangle \kappa(t)$ is invariant to the number of modes used. However, converging a calculation with 1000 bath modes to within 0.5\% (which is the error of the 1000-mode transmission coefficient in Fig~\ref{fig:modemass}) is computationally challenging, especially for low $\eta/\Delta$ where the oscillations in the transmission coefficient are very large. 

\subsection{RPMD rate}
If the alternative factorization in Eq~(\ref{eq:altfac}) is correct, and the two factors converge with the number of modes, then the product of $P(y^{\ddag})$ and $\tfrac{1}{2}\langle | \dot y | \rangle \kappa(t)$ is the RPMD rate. Calculating the transmission coefficient as the average of $\kappa(t)$ for $5 < t\Delta \leq 10$,\footnote{It does not matter greatly which time window is used for transmission coefficient calculation.} RPMD rates as a function of the number of bath modes are presented in Fig~\ref{fig:ratemode}. From this figure, one can infer that for $f \geq 20$, the overall rate is broadly invariant to the number of modes used. The deviation for $f=10$ is a systematic error, as the period of oscillations in the transmission coefficient is comparable to the length of time over which $\kappa(t)$ is averaged.

\begin{figure}[htb]
\centering
\resizebox{\figwidthf}{!} {\includegraphics[angle = 270]{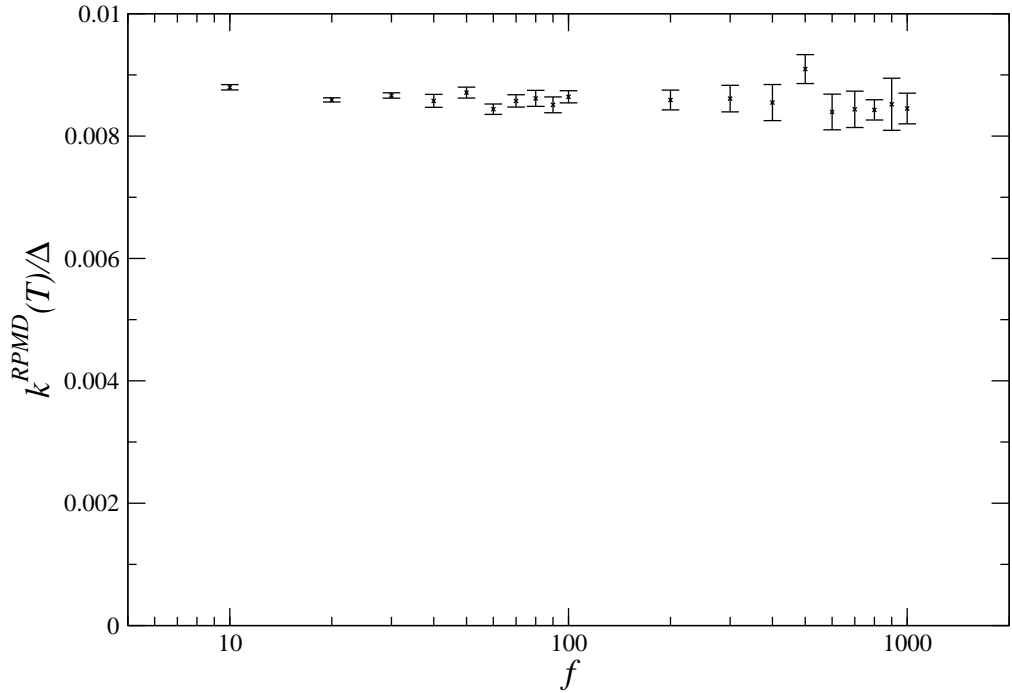}}
\caption{RPMD rate as a function of the number of bath modes $f$. The error bars are the standard deviation in $\kappa(t)$ for ten simulations of $10^5$ trajectories and include the (small) standard deviation in the accuracy of $P(y^{\ddag})$ calculations. Larger error for a greater number of bath modes is due to greater statistical error in converging a smaller transmission coefficient.}
\label{fig:ratemode}
\end{figure} 

Using this approach, RPMD rates are calculated as a function of $\eta/\Delta$ in Fig~\ref{fig:adrate}, based on the histograms in Fig~\ref{fig:transcoef}, and transmission coefficient calculations involving one bead and 1000 bath modes, where the plateau in the transmission coefficient is well-defined. Reasonable agreement is observed with the numerically exact data of Wang et.\ al.\ \cite{sb}. Note that 1000 bath modes were required in a similar semiclassical study of the spin-boson model with Debye spectral density \cite{miller_semiclas}.
\begin{figure}[htb]
\centering
\resizebox{\figwidthf}{!} {\includegraphics{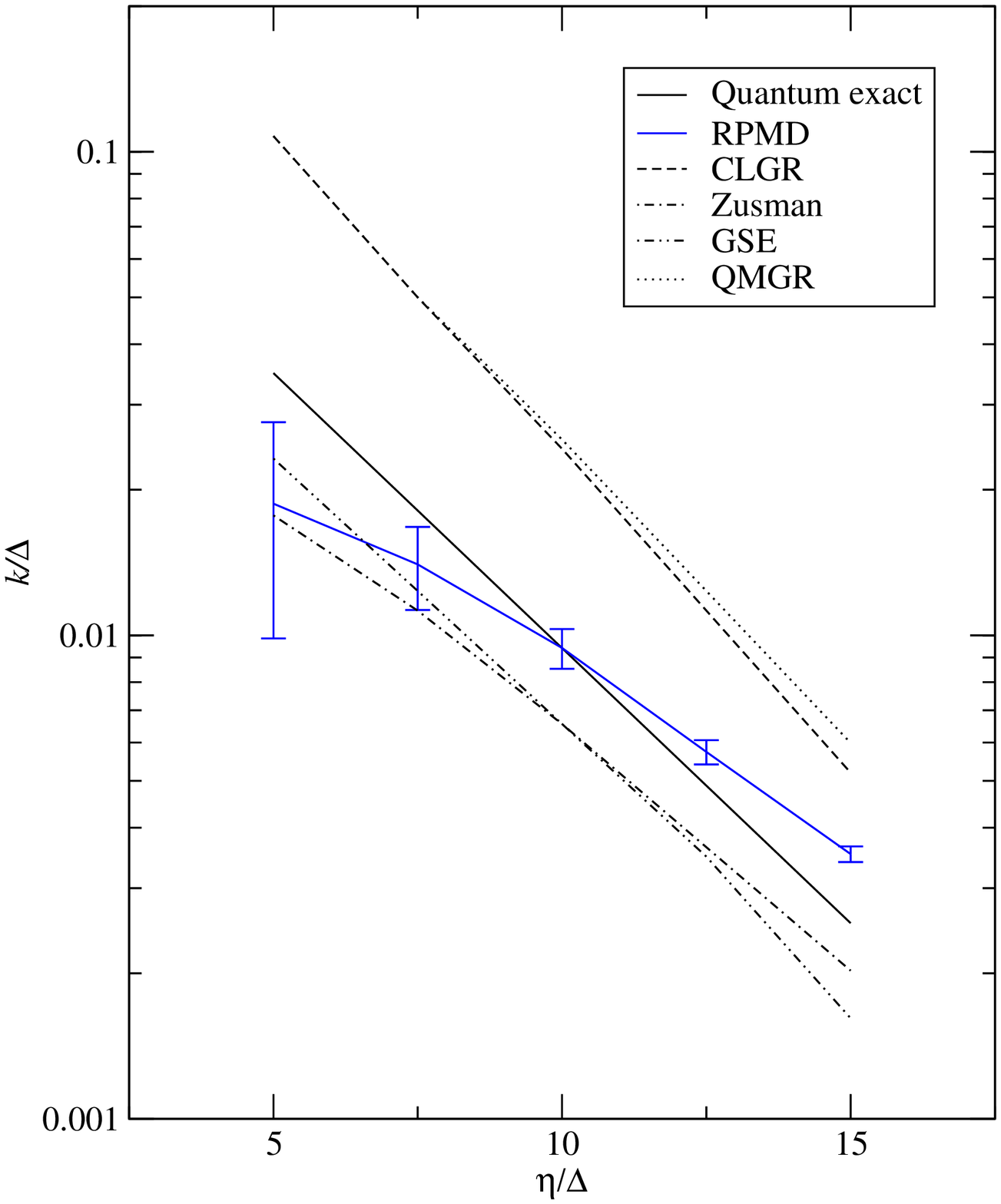}}
\caption{Reaction rates for the spin-boson model with Debye spectral density, presented as a function of the coupling strength $\eta/\Delta$ for $\Delta/\omega_c = 4$, $\beta\Delta = 0.5$. QMGR is the Quantum mechanical golden rate, CLGR the classical golden rate, GSE the generalized Smoluchowski equation and Zusman the Zusman rate. These approximate rates and the quantum exact rate were all obtained from Ref.~\cite{sb} and are included in the figure for comparison. The RPMD rate, from the calculations in this thesis, is shown in blue, and contains error bars corresponding to the standard deviation in the transmission coefficient itself, arising from the oscillations seen in Figs~\ref{fig:transcoef} and \ref{fig:modemass}, and includes the uncertainty in $P(y^{\ddag})$ calculations.}
\label{fig:adrate}
\end{figure}

The exact quantum rate appears to be underestimated by RPMD for low values of $\eta/\Delta$ and overestimated for large values of $\eta/\Delta$ such that the slope of the RPMD rate in Fig~\ref{fig:adrate} is too shallow; it is not yet apparent whether this is due to convergence or a shortcoming of RPMD rate theory. Nevertheless, the RPMD rate deviates from the exact quantum rate by less than a factor of two, and is more accurate than the other approximate methods considered in Fig~\ref{fig:adrate} for moderate coupling strengths ($7.5 \leq \eta/\Delta \leq 12.5$). The error bars show there is greater uncertainty in accuracy for low $\eta$; in the limit $\eta/\Delta \to 0$ the model tends to a system of uncoupled harmonic oscillators which does not decohere, causing large oscillations in $\kappa(t)$. 

Ref.~\cite{sb} also contained data for the $\Delta/\omega_c = 0.2$ regime; for low $\eta$ in this regime the standard deviation in $\kappa(t)$ is greater than $\kappa(t)$ itself and these results are consequently not presented.
\section{Discussion}
In this chapter I have calculated the RPMD rate in a non-adiabatic spin-boson model with Debye spectral density. The probability distribution of centroids is independent of the number of bath modes (for $f \geq 10$) and therefore independent of the mass of the reaction co-ordinate. By an alternative factorization of the rate coefficient in subsection~\ref{subsec:altfac}, the mean magnitude of the velocity of the reaction co-ordinate compensates for the vanishing transmission coefficient, leading to a convergent overall rate.

An understanding of the behaviour of the transmission coefficient can be obtained from analysis of Grote-Hynes theory \cite{gh, pollak}, which takes a very different approach to understanding the transmission coefficient. From Eq~(\ref{eq:debmass}) and as illustrated in Fig~\ref{fig:massfac}, the mass of the reaction co-ordinate vanishes in the limit of an infinite number of bath modes. As the mass falls, velocity autocorrelation dies rapidly as the ring polymer is buffeted around by a large number of bath modes. According to Grote and Hynes, the system enters a high-friction regime where the velocity `memory' of the reactant decays rapidly compared to the imaginary frequency at the top of the barrier, such that the reactant diffuses over the top of the barrier, known as the Kramers diffusion limit \cite{kramers}. In this limit, the motion of the particle is governed by Brownian forces and obeys the stochastic Langevin equation. The Grote-Hynes transmission coefficient $\kappa_{GH}$ is given by \cite{gh}
\begin{equation}
\kappa_{GH} = \lim_{t \to \infty} \kappa(t) = \frac{\mu\omega_b}{\zeta},
\end{equation}
where $\omega_b$ is the imaginary barrier frequency and $\zeta$ is the steady-state (zero-frequency) friction. Noting that $\omega_b \propto 1/\sqrt{\mu}$, we find that
\begin{equation}
 \kappa_{GH} \propto \sqrt{\mu},
\end{equation}
so as $f \to \infty$, $\mu \to 0$ and $\kappa_{GH} \to 0$, which is also observed in the long time limit of the RPMD transmission coefficients in Fig~\ref{fig:transcoef}. If we perform the unusual factorization of the rate constant seen in Eq~(\ref{eq:altfac}) such that 
\begin{equation}
 k^{RPMD}(T) = P(y^{\ddag})\times \lim_{t\to\infty} \tfrac{1}{2}\langle | \dot y | \rangle \kappa(t),
\end{equation}
and note that, from Eq~(\ref{eq:QTST}),
\begin{equation}
 \langle | \dot y | \rangle \propto \frac{1}{\sqrt{\mu}},
\end{equation}
then 
\begin{equation}
 \lim_{t\to\infty} \langle | \dot y | \rangle \kappa(t) = \langle | \dot y | \rangle \frac{\mu\omega_b}{\zeta} \propto \frac{1}{\sqrt{\mu}} \mu \frac{1}{\sqrt{\mu}} = 1.
\end{equation}
Grote-Hynes theory therefore explains why the unusual factorization avoids dependence on the vanishing mass factor\footnote{However, converging the numerical value of the rate using Grote-Hynes theory (particularly $\zeta$) is exceedingly challenging, and not presented here.}.
It is important to note that the anomalous behaviour of the mass factor is not a consequence of a non-adiabatic potential, nor of RPMD rate theory\footnote{The discretization scheme is not a function of the methodology used to compute the rate. See, for example, \ Ref \cite{markland}.}. However, it is a matter for future research whether the deviation of RPMD theory from the numerically exact quantum rate is a consequence of the inherent difficulties of the Debye spectral density, or of RPMD theory itself. 
\chapter{Conclusions}
\label{chap:c}
In this thesis, the RPMD formalism \cite{Mano_cen, Mano_pin} has been extended successfully to electronically non-adiabatic reactions, followed by application to two one-dimensional barrier transmission models and a multidimensional spin-boson model. 
After introducing RPMD rate theory in chapter~\ref{chap:rpmd}, an extension to include effects arising from electronically coupled potential energy surfaces was presented in chapter~\ref{chap:nag}. The generalisation produces physically reasonable results in the limiting case of widely separated electronic states, and in the limiting case of a single electronic surface correctly reduces to the conventional formulation of adiabatic RPMD rate theory.

Two different one-dimensional Landau-Zener models were considered in chapter~\ref{chap:1D}, with parameters chosen to accentuate non-adiabatic and tunnelling effects. Very good agreement with exact quantum results was observed across a wide range of temperatures (100--1000 K), both for a symmetric model where the dividing surface is apparent by symmetry, and an asymmetric model where the optimum dividing surface was found variationally using a QTST calculation.

A multidimensional spin-boson model with Debye spectral density was then considered in chapter~\ref{chap:sb}. For this model, the parameters were chosen such that exact quantum results and a variety of approximate calculations were available for comparison. A Bennett-Chandler factorization \cite{bc} of the RPMD rate coefficient was employed, where the rate is expressed as a product of a purely static (QTST) calculation and a dynamical (transmission coefficient) calculation. I found that the QTST rate was undefined for the Debye spectral density, and that the transmission coefficient vanished in the limit of an infinite number of bath modes. 

However, using an alternative factorization of the RPMD rate coefficient, the rate could be expressed as the product of two finite factors, such that an overall rate was obtained which was independent of the number of bath modes. This was evaluated over a range of coupling strengths, and agreement to within a factor of two was obtained compared to exact quantum results, better than almost all other approximate methods that are available for comparison. 
The unusual behaviour of the QTST rate and transmission coefficient was explained by Grote-Hynes theory \cite{gh}, which suggested that the present model is in the Kramers diffusion limit \cite{kramers}.

RPMD rate theory for electronically non-adiabatic reactions possesses numerous desirable features, including independence from the location of the dividing surface, and modelling recrossing dynamics and quantum effects. Future work could include further applications to multidimensional systems, and determining whether the deviation between exact quantum results and RPMD rate theory for the spin-boson model explored herein is a consequence of RPMD rate theory or attributable to the inherent difficulties associated with the Debye spectral density.
\appendix
\chapter{The exponential matrix}
\label{a:expm}
The requirement from chapter~\ref{chap:nag} is to calculate
 \begin{equation}
 {\bf M} (q) = e^{-\beta_n {\bf V}(q)},
\end{equation}
and 
\begin{equation}
 {\bf D}(q) = \frac{d}{dq} {\bf M}(q),
\end{equation}
where the $k$ subscripts have been dropped in this appendix for notational convenience.
The exponential matrix is defined as the convergent power series \cite{expm}
\begin{equation}
 e^{-\beta_n {\bf V}(q)} = \sum_{j=0}^{\infty} \frac{1}{j!}[-\beta_n{\bf V}(q)]^j,
\label{eq:pseries}
\end{equation}
where, for the models explored in this thesis, the potential matrix ${\bf V}(q)$ is real, symmetric and non-singular,
\begin{equation}
{\bf V}(q) = 
 \left( \begin{array}{cc}
         V_{11}(q) & V_{12} \\
	 V_{21}  & V_{22}(q)  \\ 
        \end{array}\right).
\end{equation}
Furthermore, the off-diagonal elements, $V_{12} = V_{21}$, are independent of $q$, and the diagonal elements are analytically differentiable. 

The first stage of the algorithm is to calculate the eigenvalues $\lambda_1$ and $\lambda_2$ of ${\bf V}(q)$ and their derivatives, $d \lambda_{1}/dq$ and $d \lambda_{2}/dq$. This is followed by evaluating the orthonormal matrix of eigenvectors $\bf S$ and its derivative $ d{\bf S}/dq$. Because ${\bf V}(q)$ is symmetric, it can be diagonalised by $\bf S$,
\begin{equation}
 {\bf SV }(q){\bf S^T} = {\bf \Lambda},
\end{equation}
 where ${\bf \Lambda}$ is a diagonal matrix of eigenvalues. 
The exponential of a diagonal matrix is easily calculated as a matrix of the exponential of the diagonal elements, such that
\begin{equation}
 (e^{-\beta_n{\bf \Lambda}})_{ii} = e^{-\beta_n\lambda_i},
\end{equation}
and
\begin{equation}
 \frac{d}{dq}(e^{-\beta_n{\bf \Lambda}})_{ii} = -\beta_n \frac{d\lambda_i}{dq} e^{-\beta_n\lambda_i} .
\end{equation}
From Eq~(\ref{eq:pseries}) and the orthonormality of $\bf S$ it can be shown that
\begin{eqnarray}
 e^{-\beta_n{\bf V}(q)} & = &  e^{-\beta_n {\bf S^TSV}(q){\bf S^TS}}  \\
& = & e^{{\bf S^T}(-\beta_n{\bf \Lambda) S}} \\
& = & {\bf S^T} e^{-\beta_n{\bf \Lambda}} {\bf S},
\end{eqnarray}
such that the matrix ${\bf M} (q)$ is calculated as
\begin{equation}
 {\bf M} (q) = {\bf S^T} e^{-\beta_n{\bf \Lambda}} {\bf S}.
\end{equation}
By applying the product rule to matrix differentiation, the derivative of the exponential matrix is then calculated as 
 \begin{eqnarray}
 {\bf D}(q) & = & \frac{d}{dq} {\bf M}(q) \\
 & = & \frac{d}{dq} \left( {\bf S^T}e^{-\beta_n {\bf \Lambda}}{\bf S} \right) \\
 & = & \left( \frac{d}{dq}{\bf S} \right)^{\bf T}e^{-\beta_n\bf \Lambda}{\bf S} + {\bf S^T}\left( \frac{d}{dq}e^{-\beta_n\bf \Lambda} \right) {\bf S} + {\bf S^T}e^{-\beta_n\bf \Lambda} \left( \frac{d}{dq} {\bf S} \right).
\end{eqnarray}
\chapter{Bell's Algorithm}
\label{a:bell}
To evaluate the hole matrix in Eq~(\ref{eq:hole}),
\begin{equation}
 {\bf H}_k(q_k) = {\bf M}_{k+1}(q_{k+1}) \ldots {\bf M}_n(q_n){\bf M}_1(q_1) \ldots {\bf M}_{k-1}(q_{k-1}),
\label{eq:h1}
\end{equation}
for all $k = 1, \ldots ,n$, define
\begin{equation}
 {\bf F}_k = {\bf M}_1(q_1){\bf M}_2(q_2) \ldots {\bf M}_k(q_k), 
\end{equation}
and
\begin{equation}
 {\bf G}_k = {\bf M}_k(q_k) \ldots {\bf M}_{n-1}(q_{n-1}){\bf M}_n(q_n),
\end{equation} 
such that 
\begin{equation}
 {\bf H}_k(q_k) = {\bf G}_{k+1}{\bf F}_{k-1}.
\label{eq:hcomp}
\end{equation}
Furthermore, ${\bf F}_1 = {\bf M}_1(q_1)$, ${\bf G}_n = {\bf M}_n(q_n)$ and ${\bf F}_n = {\bf G}_1 = {\bf M}_1(q_1){\bf M}_2(q_2) \ldots {\bf M}_n(q_n)$.
The algorithm proceeds in three steps:
\begin{enumerate}
 \item Set ${\bf F}_1 = {\bf M}_1(q_1)$, then compute ${\bf F}_k$, $k = 2, \ldots ,n-1$ recursively, noting that $ {\bf F}_k = {\bf F}_{k-1}{\bf M}_k(q_k)$, requiring $n-2$ matrix multiplications.
 \item Set ${\bf G}_n = {\bf M}_n(q_n)$, then compute ${\bf G}_k$, $k = n-1, n-2, \ldots, 2$ recursively, noting that $ {\bf G}_k = {\bf M}_k(q_k){\bf G}_{k+1}$, requiring $n-2$ matrix multiplications.
 \item Compute ${\bf H}_k(q_k)$ for $k=1,\ldots, n$ using Eq~(\ref{eq:hcomp}). As ${\bf H}_1(q_1) = {\bf G}_2$ and ${\bf H}_n(q_n) = {\bf F}_{n-1}$, this only requires $n-2$ matrix multiplications.
\end{enumerate}
This algorithm\footnote{This is attributed to Martin Bell, after whom it is named.} computes all hole matrices in $3n-6$ matrix multiplications, i.e.\ is $O(n)$. Conversely, evaluating each ${\bf H}_k(q_k)$ independently using Eq~(\ref{eq:h1}) requires $n(n-2)$ matrix multiplications, so is $O(n^2)$, leading to far greater computational effort for large $n$.
\chapter{Mass Factor Identity}
\label{a:massfac}
From Eq~(\ref{eq:massfac}) on p.~\pageref{eq:massfac}, the identity to be proven is
\begin{equation}
 \sum_{k=1}^f\tan^2\left(\frac{(k-\tfrac{1}{2})\pi}{2f}\right) = f(2f-1).
\end{equation}
Noting that $\tan^2\theta \equiv \sec^2 \theta - 1$, this identity can equivalently be stated as
\begin{equation}
 \sum_{k=1}^f \sec^2 \theta_k = 2f^2,
\label{eq:sec}
\end{equation}
where, for brevity, I have defined 
\begin{equation}
 \theta_k =  \frac{(k-\tfrac{1}{2})\pi}{2f}.
\end{equation}
Equation (\ref{eq:sec}) is proven by analysing the coefficients of a polynomial whose roots are $\sec^2\theta_k$. Consider the polynomial
\begin{equation}
 \prod_{k=1}^f (x - \sec^2 \theta_k) = \sum_{j=0}^f c_jx^j.
\label{eq:poly}
\end{equation}
By expanding the left hand side of Eq~(\ref{eq:poly}), we see that the term corresponding to $x^{f-1}$ is the negative sum of the roots, 
\begin{equation}
 c_{f-1} = -\sum_{k=1}^f \sec^2\theta_k,
\label{eq:coeff}
\end{equation}
and as such, I need to prove that 
\begin{equation}
 c_{f-1} = -2f^2.
\end{equation}
Substituting $y^2 = 1/x$ into the left hand side of Eq~(\ref{eq:poly}), and rearranging,
\begin{eqnarray}
 \prod_{k=1}^f (x - \sec^2 \theta_k) & = & \left(\prod_{k=1}^f\sec^2 \theta_k\right)^{-1} \prod_{k=1}^f (x\cos^2\theta_k - 1) \nonumber \\
& = & N y^{-2f} \prod_{k=1}^f (\cos^2\theta_k - y^2) \nonumber \\
& = & N y^{-2f} \prod_{k=1}^f (\cos\theta_k - y)(\cos\theta_k + y)\label{eq:step1}\\
& = & N y^{-2f} (-1)^f \prod_{k=1}^{2f} (y-\cos\theta_k) \label{eq:step2} 
\label{eq:ppowers}.
\end{eqnarray}
The inverse product of the roots is
\begin{equation}
 N = \left(\prod_{k=1}^f\sec^2 \theta_k\right)^{-1},
\end{equation}
and I have used the periodic properties of the cosine function to move from Eq~(\ref{eq:step1}) to Eq~(\ref{eq:step2}), as $\cos\theta_k = -\cos\theta_{2f-k}$. 
Equation~(\ref{eq:step2}) leads to a new polynomial,
\begin{equation}
(-1)^f \prod_{k=1}^{2f} (y-\cos\theta_k) = \sum_{l=0}^{2f} a_l y^{l},
\label{eq:newpoly}
\end{equation}
where 
\begin{eqnarray}
 a_0 & = & (-1)^f \prod_{k=1}^{2f} -\cos\theta_k \nonumber \\
& = & \prod_{k=1}^{f} \cos^2\theta_k \nonumber \\
& = & N^{-1}.
\end{eqnarray}
However, from Eqs~(\ref{eq:poly}), (\ref{eq:ppowers}) and (\ref{eq:newpoly}), $N a_{2(f-j)} = c_j$, so
\begin{equation}
 c_{f-1} = N a_2 = \frac{a_2}{a_0}.
\label{eq:apower}
\end{equation}
Examining Eq~(\ref{eq:newpoly}), the task is now to find a polynomial with known coefficients, whose roots are $\cos\theta_k$. To do this, note that
\begin{equation}
 \cos(2f\theta_k) = \cos[(k-\tfrac{1}{2})\pi] = 0,
\end{equation}
for all integer values of $k$. If we write $\cos(2f\phi)$ as a polynomial in powers of $\cos\phi$, the polynomial will have roots when $\cos \phi = \cos\theta_k$, and we can therefore substitute $y = \cos\phi$ into Eq~(\ref{eq:newpoly}). Expanding $\cos(2f\phi)$ using De Moivre's theorem,
\begin{eqnarray}
 \cos(2f\phi) & = & \mathfrak{R}\ e^{2fi\phi} \nonumber \\
& = & \mathfrak{R}\ (\cos\phi + i\sin\phi)^{2f} \nonumber \\
& = & \mathfrak{R}\ \sum_{j=0}^{2f} \binom{2f}{j} \cos^{2f-j}\phi \ i^j \sin^j\phi \nonumber \\
& = & \sum_{k=0}^{f}\binom{2f}{2k} \cos^{2(f-k)}\phi\ (-1)^k \sin^{2k}\phi \nonumber \\
& = & \sum_{k=0}^f \binom{2f}{2k} \cos^{2(f-k)}\phi\ (-1)^k \sum_{l=0}^{k}\binom{k}{l} (-1)^l \cos^{2l}\phi \label{eq:cos2mf} \\
& = & (-1)^f  \sum_{l=0}^{2f} a_l y^{l},
\end{eqnarray}
where 
\begin{equation}
 \binom{k}{l} = \frac{k!}{l!(k-l)!},
\end{equation}
are binomial coefficients, and $\mathfrak{R}$ denotes the real part. From Eq~(\ref{eq:apower}), the ratio of $a_2$ and $a_0$ are required. The part of the expansion of Eq~(\ref{eq:cos2mf}) which is $O(\cos^0\phi)$ occurs when $k = f$, $l=0$, such that
\begin{equation}
 a_0 = (-1)^f \binom{2f}{2f} (-1)^f \times \binom{f}{0} = 1. 
\end{equation}
The terms of Eq~(\ref{eq:cos2mf}) which are $O(\cos^2\phi)$ have $k = f - 1$ and $l = 0$, or $k = f$ and $l = 1$, such that
\begin{eqnarray}
a_2 & = & (-1)^f\binom{2f}{2f-2} (-1)^{f-1} \times \binom{f-1}{0} + (-1)^f\binom{2f}{2f} (-1)^{f+1} \binom{f}{1}  \\
& = & -(2f^2 -f) - f \\
& = & -2f^2. \label{eq:a2}
\end{eqnarray}
 Combining the results of Eqs~(\ref{eq:coeff}), (\ref{eq:apower}) and (\ref{eq:a2}),
\begin{eqnarray}
 \sum_{k=1}^f \sec^2\theta_k & = & -c_{f-1} \nonumber \\
& = & -\frac{a_2}{a_0}\nonumber \\
& = & -\frac{-2f^2}{1} \nonumber \\
& = & 2f^2,
\end{eqnarray}
as required\footnote{I would like to acknowledge assistance from Dr N.\ J.\ B.\ Green with the formulation of this proof.}.
\addcontentsline{toc}{chapter}{References}
\bibliography{refs1}
\end{document}